\documentclass{aa}  
\usepackage{graphicx}
\usepackage[varg]{txfonts}
\usepackage{subfig}
\usepackage{newtxtext,newtxmath}
\usepackage[labelfont=bf]{caption}
\usepackage{subfig}
\usepackage[table]{xcolor}

\numberwithin{equation}{section} 
\newcommand{\kms}{~km\,s$^{-1}$}
\graphicspath{{images_final/}}
\newcommand{\msun}{M$_\odot$}

\usepackage{hyperref}
\hypersetup{colorlinks=true,linkcolor=red,citecolor=blue,urlcolor=magenta}

\defcitealias{2019Bacchini}{B19a}
\defcitealias{2019TheBook}{CFN19}

\begin{document} 
    \title{Evidence for supernova feedback sustaining gas turbulence in nearby star-forming galaxies}
   \authorrunning{C. Bacchini et al.}
   \author{Cecilia Bacchini\inst{1,2,3},
          Filippo Fraternali\inst{1},
          Giuliano Iorio\inst{4},
          Gabriele Pezzulli\inst{5}, 
          Antonino Marasco\inst{6}, \and
          Carlo Nipoti\inst{2}          
          }

   \institute{Kapteyn Astronomical Institute, University of Groningen, Landleven 12, 9747 AD Groningen, The Netherlands\\
   \email{c.bacchini@rug.nl}
   \and
   Dipartimento di Fisica e Astronomia, Universit\`{a} di Bologna, via Gobetti 93/2, I-40129, Bologna, Italy\\
   \email{cecilia.bacchini@unibo.it}
   \and
   INAF - Osservatorio di Astrofisica e Scienza dello Spazio di Bologna, via Gobetti 93/3, I-40129, Bologna, Italy
   \and
   Dipartimento di Fisica e Astronomia, Universit\`{a} di Padova, Vicolo dell’Osservatorio 3, IT-35122, Padova, Italy
   \and
   Department of Physics, ETH Zurich, Wolfgang-Pauli-Strasse 27, 8093 Zurich, Switzerland
   \and
   INAF - Osservatorio Astrofisco di Arcetri, largo E. Fermi 5, 50127 Firenze, Italy
   }

   \date{Received ; accepted .}

 
  \abstract{
  It is well known that gas in galaxy discs is highly turbulent, but there is much debate on which mechanism can energetically maintain this turbulence. 
  Among the possible candidates, supernova (SN) explosions are likely the primary drivers but doubts remain on whether they can be sufficient in regions of moderate star formation activity, 
  in particular in the outer parts of discs. 
  Thus, a number of alternative mechanisms have been proposed. 
  In this paper, we measure the SN efficiency $\eta$, namely the fraction of the total SN energy needed to sustain turbulence in galaxies, and verify that SNe can indeed be the sole driving 
  mechanism. 
  The key novelty of our approach is that we take into account the increased turbulence dissipation timescale associated to the flaring in outer regions of gaseous discs. 
  We analyse the distribution and kinematics of HI and CO in 10 nearby star-forming galaxies to obtain the radial profiles of the kinetic energy per unit area, for both the atomic gas and 
  the molecular gas. 
  We use a theoretical model to reproduce the observed energy with the sum of turbulent energy from SNe, as inferred from the observed star formation rate (SFR) surface density, 
  and the gas thermal energy. 
  For the atomic gas, we explore two extreme cases in which the atomic gas is made either of cold neutral medium or warm neutral medium, and the more realistic scenario with a mixture of 
  the two phases. 
  We find that the observed kinetic energy is remarkably well reproduced by our model across the whole extent of the galactic discs, assuming $\eta$ constant with the galactocentric radius. 
  Taking into account the uncertainties on the SFR surface density and on the atomic gas phase, we obtain that the median SN efficiencies for our sample of galaxies are
  $\langle \eta_\mathrm{atom} \rangle = 0.015_{-0.008}^{+0.018}$ for the atomic gas and $\langle \eta_\mathrm{mol} \rangle = 0.003_{-0.002}^{+0.006}$ for the molecular gas. 
  We conclude that SNe alone can sustain gas turbulence in nearby galaxies with only few percent of their energy and that there is essentially no need for any further source of energy. 
  }


   \maketitle

\section{Introduction}\label{sec:intro}
Gas kinematics provides valuable information about the physical properties of the interstellar medium (ISM). 
In particular, the velocity dispersion measured from the broadening of emission lines is fundamental to study turbulence. 
Several authors have analised the kinematics of atomic and molecular gas in nearby star-forming galaxies, finding that the velocity dispersion 
shows a decreasing trend with the galactocentric radius \citep[e.g.][]{2002Fraternali,2008Boomsma,2011Wilson,2016Mogotsi,2017Iorio,2019Bacchini}.
In the inner regions of galaxies, the gas velocity dispersion is typically about 15--20 \kms, well exceeding the expected broadening due to thermal motions alone 
(i.e. $\lesssim 8$ \kms). 
This non-thermal broadening is usually ascribed to turbulence. 
At large galactocentric radii instead (where the molecular gas emission is typically not detected), the velocity dispersion of HI approaches values that are compatible with the thermal 
broadening of the warm neutral gas (at temperature $T\approx 8000$~K). 

Both thermal and turbulent motions are forms of disordered energy, but while the first is related to the temperature of the gas particles, the second can be seen as 
the relative velocity between macroscopic portions of the fluid.   
The behaviour of turbulence in incompressible fluids is well described by Kolmogorov's theory \citep{1941Kolmogorov}, which we briefly outline in the following (see e.g. 
\citealt{2004ElmegreenScalo} for details). 
Turbulence entities are envisioned as ``eddies'' that develop at a variety of different spatial scales. 
Turbulent energy is injected on a certain scale $L_\mathrm{D}$, called driving scale, at which the largest eddies are formed. 
The largest eddies break down into smaller and smaller eddies and transfer kinetic energy to smaller scales in the so-called ``turbulent cascade'', until the dissipation scale 
$l_\mathrm{d}$ is reached. 
The energy is conserved throughout this cascade for any scale between the driving scale and the dissipation scale (i.e. inertial range). 
At the dissipation scale instead, viscosity transforms the turbulent kinetic energy into internal energy. 
Kolmogorov's framework is usually assumed to describe the ISM turbulence, despite the gas in galaxies is expected to be compressible in the presence of supersonic turbulent motions  
\citep[e.g.][]{2004ElmegreenScalo}. 
There are indeed observational indications that Kolmogorov's theory might be adequate to model ISM turbulence \citep[e.g.][]{2001Elmegreen,2009Dutta}. 

Both Kolmogorov's theory and numerical simulations of compressible supersonic turbulence in the ISM show that the turbulent energy should be rapidly dissipated on timescales of 
the order of 10~Myr \citep[e.g.][]{1998Stone,1998MacLow,1999Padoan,1999MacLow}. 
Hence, a continuous source of energy is needed in order to maintain the turbulence of the gas ubiquitously observed in galaxies. 
This issue has stimulated previous research to understand which mechanism is feeding turbulence in galaxies \citep[e.g.][]{2009Tamburro,2010Klessen,2013Stilp,2019Utomo}. 
Among the possible candidates, there are different forms of stellar feedback, which include proto-stellar jets, winds from massive stars, ionising radiation, and supernovae (SNe). 
These latter likely dominate the energy input with respect to the other mechanisms \citep[][]{2004MacLowKlessen,2004ElmegreenScalo}. 
SN explosions are extremely powerful phenomena that can inject a huge amount of energy into the ISM, even though most of this energy is expected to be radiated away \citep[e.g.][]{1977McKee}. 
Numerical simulation of SN remnant evolution in the ISM have consistently shown that the efficiency of SNe, which is typically defined as the fraction of the total SN energy that is 
injected into the ISM as kinetic energy, is $\sim 0.1$ \citep[e.g.][but see also \citealt{2018Fielding}]{1998Thornton,2005Dib,2015bKim,2016Martizzi,2016Fierlinger,2019Ohlin}. 
This appears at odds with a number of recent works showing that the observed kinetic energy of the atomic gas in nearby galaxies requires a SN feedback with a efficiency of 
$\approx 0.8-1$ \citep[e.g.][]{2009Tamburro,2013Stilp,2019Utomo}.  
Therefore, other physical mechanisms have been considered as additional drivers of turbulence, like magneto-rotational instability \citep[MRI, e.g.][]{1999Sellwood}, gravitational 
instability \citep[e.g.][]{2016Krumholz}, rotational shear \citep[e.g.][]{2002Wada}, and accretion flows \citep[e.g.][]{2010Klessen,2010Krumholz,2010Elmegreen}. 
However, quantifying the amount of kinetic energy provided by these mechanisms is not straightforward and the values predicted by the models are affected by large 
uncertainties on the observable quantities (e.g. mass accretion rate, magnetic field intensity). 
Hence, it is still not clear which (if any) of these additional sources of energy are at play. 

A further difficulty in studying the turbulent energy is represented by the challenge of disentangling thermal and turbulent motions in observations. 
This issue is particularly significant for the atomic gas, which is expected to be present in two phases with different temperatures: the cold neutral medium (CNM) with 
$T\approx 80$~K, and the warm neutral medium (WNM) with $T\approx 8000$~K \citep{1995Wolfire,2003Wolfire}. 
This latter can significantly contribute to the velocity dispersion ($\lesssim 8$ \kms) and the kinetic energy, but the lack of information about the relative fraction of CNM and WNM 
is usually an irksome obstacle to interpreting the observed velocity dispersion. 
In the Milky Way, \cite{2003Heiles} estimated that approximately 60\% of the atomic hydrogen in the solar neighborhood is WNM (at latitudes larger than 10\degr).
\cite{2013Pineda} found that the fraction of WNM is $\sim$30\% and $\sim$80\% within and beyond the solar radius respectively. 
However, there are indications of significant variations between galaxies, preventing from adopting the Galactic values for extra-galactic studies. 
Indeed, \cite{1993Dickey} measured that the WNM represents $\sim$60\% and $\sim$85\% of the total HI in M31 and M33, respectively. 
In the Large Magellanic Cloud, \cite{2000MarxZimmer} found that the WNM is about 65\%, while \cite{2000Dickey} estimated a lower limit of $\sim$85\% for the Small Magellanic 
Cloud \citep[see also][]{2019Jameson}.  
High fractions of WNM were also claimed by \cite{2012Warren}, who studied HI line profiles for a sample of 27 nearby galaxies and found that, despite the CNM phase is present 
in almost all their galaxies, it is only a few percent of the total HI. 

The purpose of this work is to understand whether SNe can provide sufficient energy to maintain the turbulence of neutral gas in nearby star-forming galaxies and, in particular, 
to infer the SN efficiency. 
The main improvement with respect to previous works is that we take into account the radial flaring of gas discs in galaxies, which implies longer timescales of turbulence dissipation. 
Moreover, we use a Bayesian method to effectively explore the parameter space of our model of SN-driven turbulence. 
In this paper, Section~\ref{sec:obs} describes the sample of galaxies and the observations used to measure the kinetic energy. 
In Sec.~\ref{sec:method}, we explain how we derive the turbulent energy and the thermal energy components expected from ISM models, and the method used to obtain the SN efficiency 
from a set of observations. 
In Sec.~\ref{sec:results}, we show the resulting profiles of the energy components and provide the SN efficiencies for the galaxies in our sample. 
In Sec.~\ref{sec:discussion}, we derive a ``global'' value for efficiency of the atomic and the molecular gas by considering the whole sample of galaxies; we then discuss our 
findings in the broader context of self-regulating star formation and compare our results with previous works in the literature on SN feedback and other driving mechanisms. 
Section~\ref{sec:conclusions} summarises this work and draws the main conclusions. 

\section{Observations and galaxy sample}\label{sec:obs}
Given suitable emission-line spectroscopic observations, the kinetic energy per unit area of the (atomic or molecular) gas in a galaxy as a function of the galactocentric radius 
$R$ can be estimated as
\begin{equation}\label{eq:E_obs_def}
\begin{split}
 E_\mathrm{obs}(R)	& = \frac{3}{2} \Sigma(R) \sigma^2(R) \\
			& \simeq \left (3 \times 10^{46} \text{ erg} \, \text{pc}^{-2} \right) 
			\left( \frac{\Sigma}{10 \text{ \msun pc}^{-2}} \right) 
			\left( \frac{\sigma}{10 \text{ \kms}} \right)^2 \, ,
\end{split}
\end{equation}
where $\Sigma(R)$ is the surface density, $\sigma(R)$ is the velocity dispersion, and the factor of 3 in the first equality comes from the assumption of isotropic velocity dispersion. 
We calculated Eq.~\ref{eq:E_obs_def} for the neutral gas in a sample of nearby star-forming galaxies by studying the distribution and kinematics of HI and CO, which is adopted as H$_2$ 
tracer. 
All the radial profiles used in this work are azimuthal averages calculated by dividing the galaxy into concentric tilted rings of about 400~pc width 
\citep[see][hereafter \citetalias{2019Bacchini}]{2019Bacchini}. 

\subsection{Atomic gas distribution and kinematics}\label{sec:obs_HI}
In this work, we used the velocity dispersion derived in \citetalias{2019Bacchini}, in which we studied the HI kinematics in 12 nearby star-forming galaxies using 21-cm emission line data 
cubes from The HI Nearby Galaxy Survey \citep[THINGS;][]{2008Walter}. 
The data cubes were analysed using the software $^{\text{\textsc{3D}}}$\textsc{Barolo} \citep{2015Diteodoro}, which carries out a tilted-ring model fitting on emission-line data cubes. 
$^{\text{\textsc{3D}}}$\textsc{Barolo} can take into account the beam smearing effect and robustly measure the velocity dispersion and the rotation curve of a galaxy, performing significantly 
better than 2D methods based on moment maps \citep[e.g.][]{2016DiTeodoro,2017Iorio}. 
Figure 2 in \citetalias{2019Bacchini} shows the radial profiles of the HI velocity dispersion ($\sigma_\mathrm{HI}$) for the sample: typically, 
$\sigma_\mathrm{HI}$ is 15--20 \kms in the inner regions of galaxies and 6--8 \kms in the outskirts. 
For NGC~6946, we performed a new kinematic analysis using the 21-cm data cube in \cite{2008Boomsma} with spatial resolution of 13\arcsec (i.e. $\approx 330$~pc), 
as it has an higher signal-to-noise ratio with respect to the THINGS data cube used in \citetalias{2019Bacchini}. 

For highly inclined or warped galaxies, the line of sight intercepts regions with different rotation velocity, which can artificially broaden the line profile. 
This issue biases the velocity dispersion towards high values and $^{\text{\textsc{3D}}}$\textsc{Barolo} cannot correct for this effect. 
Hence, we decided to exclude two galaxies from this study, NGC~2841 ($i\approx 74$\degr) and NGC~7331 ($i\approx 76$\degr), as their average HI velocity 
dispersion is systematically $\gtrsim 5$ \kms above that of the other galaxies. 
NGC~3198 instead, despite the relatively high inclination ($i\approx 72$\degr), appears much less affected by this issue and is then included in our sample (see Appendix D in 
\citetalias{2019Bacchini} for a more detailed discussion). 
In addition, NGC~5055 shows a warp along the line of sight, which starts beyond $R\approx 10$~kpc \citep{2006Battaglia}. 
In these regions, the velocity dispersion is systematically higher than the typical values in the outer parts of star-forming galaxies, as the line profile is broadened by the merging of 
emission from different annuli intercepted by the line of sight. 
Hence, we excluded these regions from this study. 
We obtain a final sample of 8 spiral galaxies and 2 dwarf galaxies (i.e. DDO~154 and IC~2574).

We derived the HI surface density ($\Sigma_\mathrm{HI}$) as a function of the galactocentric radius with the task \texttt{ELLPROF} of $^{\text{\textsc{3D}}}$\textsc{Barolo}, 
which provides the radial profiles corrected for the galaxy inclination. 
For NGC~0925, NGC~2403, NGC~3198, and NGC~5055, we used the publicly available data cubes from the Hydrogen Accretion in LOcal GAlaxieS (HALOGAS) Survey \citep{2011Heald}, which have a 
better signal-to-noise ratio with respect to the robust-weighted THINGS data cubes adopted in \citetalias{2019Bacchini}. 
DDO~154, IC~2574, NGC~2976, NGC~4736, and NGC~7793 are not included in the HALOGAS sample, hence we obtained $\Sigma_\mathrm{HI}$ from the natural-weighted THINGS data cubes, as they have a 
better signal-to-noise ratio with respect to the robust-weighted data cubes. 
In the case of NGC~6946, we employed the same data cube as in \cite{2008Boomsma}. 
We obtained the surface density of the total atomic gas by accounting for the Helium fraction (ie. $\Sigma_\mathrm{atom} = 1.36 \Sigma_\mathrm{HI}$). 
We also verified that our profiles are compatible with previous estimates in the literature \citep{2002Fraternali,2006Battaglia,2008Leroy,2008Boomsma,2010Bigiel,2013Gentile,2017Iorio}. 

\subsection{Molecular gas distribution and kinematics}\label{sec:obs_CO}
We measured the velocity dispersion of CO, the typical tracer of the molecular gas, using $^{\text{\textsc{3D}}}$\textsc{Barolo} on CO(2-1) emission line data cubes from the HERA 
CO-Line Extragalactic Survey \citep[HERACLES;][]{2005Leroy}. 
The emission is detected in 7 out of 10 galaxies of our sample, except DDO~154, IC~2547, and NGC~7793. 
We provide the results of the kinematic analysis (e.g. moments maps, position-velocity diagrams, rotation curves) in Appendix~\ref{ap:3DB}. 
In \citetalias{2019Bacchini}, we did not analyse the molecular gas kinematics for each galaxy in the sample, as we relied on previous works in the literature that showed that the velocity 
dispersion of CO is about half of $\sigma_\mathrm{HI}$ \citep[][]{2016Mogotsi,2017Marasco,2019Koch}. 
For this work, a robust measurement of the velocity dispersion is desirable to accurately calculate Eq.~\ref{eq:E_obs_def}, allowing also to test the assumption used in 
\citetalias{2019Bacchini} (see Appendix~\ref{ap:3DB}). 
We indicate the molecular gas velocity dispersion with $\sigma_\mathrm{H_2}$. 
We must note however that it is not clear whether the nature of the non-thermal component of the molecular gas velocity dispersion is dynamic (i.e. disordered motions between 
self-gravitating clouds) and hydro-dynamic (i.e. disordered motions between portions of fluid, similarly to the atomic gas turbulence). 
Our approach based on Kolmogorov's theory adheres to the second scenario. 

We took the radial profiles of the molecular gas surface density ($\Sigma_\mathrm{mol}$) from \cite{2016Frank}. 
These authors measured the CO luminosity using HERACLES data cubes and derived $\Sigma_\mathrm{mol}$ using the CO-to-H$_2$ conversion factor from \cite{2013Sandstrom}. 
These latter authors obtained the radial profile of the conversion factor in a sample of 26 galaxies taking into account the dust-to-gas ratio and the metallicity gradient. 
NGC~2403 was not included in \cite{2013Sandstrom} study, hence \cite{2016Frank} adopted the MW value for the conversion factor. 
The profiles of \cite{2016Frank}, which already include the Helium correction, are shown in Fig.~1 in \citetalias{2019Bacchini}, where the errorbars take into account the uncertainty 
on the CO-to-H$_2$ conversion factor.

\subsection{Star formation rate surface density}\label{sec:obs_SFRD}
To estimate the turbulent energy produced by SNe, we used the observed SFR surface density ($\Sigma_\mathrm{SFR}$).
We took as references two different estimates of the SFR surface density, one from \cite{2008Leroy} and \cite{2010Bigiel}, and the other from \cite{2009MunozMateos}. 
This allows us to test the possible dependence of our results on different methods to derive the SFR surface density from the observations. 

The profiles from \cite{2008Leroy} are obtained by combining the far-ultraviolet (FUV, i.e. unobscured SF) emission maps from the Galaxy Evolution 
Explorer \citep[GALEX;][]{2007GildePaz} and the 24~$\mu$m (obscured SF) emission maps from the Spitzer Infrared Nearby Galaxy Survey \citep[SINGS;][]{2003Kennicutt}. 
We also included the profiles from \cite{2010Bigiel}, which are derived from FUV GALEX maps out to larger radii with respect to \cite{2008Leroy}. 

For DDO~154, NGC~2403, NGC~3198, and NGC~6946, $\Sigma_\mathrm{SFR}(R)$ is less radially extended than $\Sigma_\mathrm{HI}(R)$, as the FUV emission from the outermost radii 
goes below the sensitivity limit of the observations. 
In particular, 9 out of 10 galaxies are in the sample of \cite{2010Bigiel}, hence the upper limit on $\Sigma_\mathrm{SFR}(R)$ is $2 \times 10^{-5}$~M$_\odot$yr$^{-1}$kpc$^{-2}$. 
For NGC~6946, which is not included in that study, the upper limit is $10^{-4}$~M$_\odot$yr$^{-1}$kpc$^{-2}$ \citep{2008Leroy}. 
In our modelling procedure (see Sec.~\ref{sec:method}), we use these upper limits to constrain the energy injected by SNe at large radii, rather than simply discarding the 
outskirts of galaxies from our analysis. 

To derive a radial profile of SFR surface density from the data of \cite{2009MunozMateos}, we followed the same procedure described in \cite{2015Pezzulli}. 
We adopted the UV extinction radial profiles based on the dust attenuation prescription from \cite{2008Cortese}.
Among our galaxies, only the dwarf galaxy DDO~154 is not included in this sample. 
In general, the values of the SFR surface densities derived from \cite{2009MunozMateos} are above those from \cite{2008Leroy}, in particular at large radii, while $\Sigma_\mathrm{SFR}(R)$ 
in the innermost regions of NGC~0925, NGC~3198, and NGC~6946 is reduced. 
We anticipate that our main conclusions do not change whether we use one or the other determination of the SFR surface density. 


\section{Methods}\label{sec:method}
In this section, we first describe the energy components that we took into account to estimate theoretically the energy of the atomic gas and the molecular gas. 
We then explain the method used to compare this energy with the energy profiles calculated from the observations using Eq.~\ref{eq:E_obs_def} in order to obtain the SN efficiency. 

\subsection{Energy components}\label{sec:method_components}
We assume that the total energy of the (atomic or molecular) gas per unit area ($E_\mathrm{mod}$) is the sum of the turbulent energy ($E_\mathrm{turb}$) and 
the thermal energy ($E_\mathrm{th}$) 
\begin{equation}\label{eq:E_mod_def}
 E_\mathrm{mod}(R) = E_\mathrm{turb}(R) + E_\mathrm{th}(R) \, .
\end{equation}
Hence, the velocity dispersion of the gas is
\begin{equation}\label{eq:sigma_obs_def}
\sigma_\mathrm{mod}(R)=\sqrt{ \upsilon_\mathrm{turb}^2(R) + \upsilon_\mathrm{th}^2(R) } \, ,
\end{equation}
where $\upsilon_\mathrm{turb}$ is the turbulent velocity and $\upsilon_\mathrm{th}$ is the thermal velocity. 
The equation for the turbulent energy is described below in Sec.~\ref{sec:method_Eturb} and is the same for the atomic gas and the molecular gas. 
For the thermal component instead, we discriminate between the atomic and the molecular gas (see Sec.~\ref{sec:method_Eth}). 

\subsubsection{Turbulent energy from supernova feedback}\label{sec:method_Eturb}
The timescale of turbulence dissipation is defined as the ratio between the turbulent energy and its dissipation rate \citep[e.g.][]{2004ElmegreenScalo,2004MacLowKlessen}. 
In the stationary Kolmogorov's regime, the dissipation rate at the viscosity scale must be equal to the injection rate at the driving scale $\dot{E}_\mathrm{turb}$. 
Therefore, the dissipation timescale can be written as \citep[e.g.][]{1999MacLow}
\begin{equation}\label{eq:tau_d_def}
 \tau_\mathrm{d} \equiv \frac{E_\mathrm{turb}}{\dot{E}_\mathrm{turb}} 
		  = \frac{L_\mathrm{D}}{\upsilon_\mathrm{turb}} 
		  \simeq \left( 10 \text{ Myr} \right)
		  \left( \frac{L_\mathrm{D}}{100 \text{ pc}} \right)
		  \left( \frac{\upsilon_\mathrm{turb}}{10 \text{\kms}} \right) ^{-1} \, ,
\end{equation}
which corresponds to the crossing time of turbulent gas across the driving scale $L_\mathrm{D}$ \citep[e.g.][]{2000Elmegreen}. 
This latter is difficult to measure precisely from observations and likely depends on the size of the physical system and the mechanisms under consideration. 
We note that, as Eq.~\ref{eq:E_mod_def} and Eq.~\ref{eq:sigma_obs_def}, Eq.~\ref{eq:tau_d_def} is valid in general for any source of turbulent energy. 

This work is focused on SNe, as they are expected to dominate the energy input by stellar feedback on the scale of galactic discs (see Sec.~\ref{sec:discussion_othersources} 
for a discussion on other possible sources). 
We therefore adopted a specific equation to calculate $E_\mathrm{turb}$ in Eq.~\ref{eq:E_mod_def} in the case of SNe (i.e. $E_\mathrm{turb,SNe}$). 
In particular, we assumed that 
\begin{equation}\label{eq:Ld_2h}
L_\mathrm{D} = 2 h\, , 
\end{equation}
where $h$ is the scale height of the gas disc, and it is $h_\mathrm{HI}$ for the atomic gas and $h_\mathrm{H_2}$ for the molecular gas. 
This choice is motivated by theoretical models of SN remnant evolution and by observational evidence showing that the explosion of multiple SNe generates expanding shells (super-bubbles), 
whose diameter can easily reach the thickness of the atomic gas disc \citep[e.g.][]{1989MacLow,2008Boomsma}. 
In addition, $L_\mathrm{D}$ corresponds to the size of the largest eddies, which are expected to be approximately as large as the physical scale of the system. 
We discuss in depth the assumption of Eq.~\ref{eq:Ld_2h} in Sec.~\ref{sec:discussion_caveats}. 

The gas scale height for each galaxy in our sample was calculated with the same method as in \citetalias{2019Bacchini}: the gas is assumed in vertical hydrostatic equilibrium in the galactic 
potential and $h$ is derived iteratively with the Python module \textsc{Galpynamics}\footnote{\url{https://github.com/iogiul/galpynamics}} 
\citep{2018Iorio} in order to also take into account the gas self-gravity. 
The scale height of the gas distribution increases for increasing velocity dispersion $\sigma$ and for decreasing intensity of the vertical gravitational force $g_z$. 
Both $g_z$ and $\sigma$ decrease with radius, but the first effect is dominant. 
Hence, the gas distribution flares with the radius and the scale height increases, reaching hundreds of parsecs in the outer regions of discs. 
For each galaxy, $h_\mathrm{HI}$ was calculated for the gravitational potential produced by stars and dark matter (see \citetalias{2019Bacchini} for details), 
and then $h_\mathrm{H_2}$ was derived including also the potential of the atomic gas distribution with the flare. 
\footnote{This choice implies that the atomic gas distribution is not influenced by the molecular gas distribution. 
We expect that including this latter does not significantly affect $h_\mathrm{HI}$, as the molecular gas is concentrated in the inner regions of the galaxies, where stars are the dominant mass component.}
The radial profiles of $h_\mathrm{HI}$ for our galaxies can be found in Fig. 4 in \citetalias{2019Bacchini}. 
A major improvement of this work is that $h_\mathrm{H_2}$ was calculated using the velocity dispersion measured from CO data (see Sec.~\ref{sec:results_H2}). 
We, however, found that the resulting profiles are compatible within the errors with those obtained in \citetalias{2019Bacchini}, where we assumed $\sigma_\mathrm{H_2} 
\approx \sigma_\mathrm{HI}/2$. 

\begin{figure}
\includegraphics[width=1.\columnwidth]{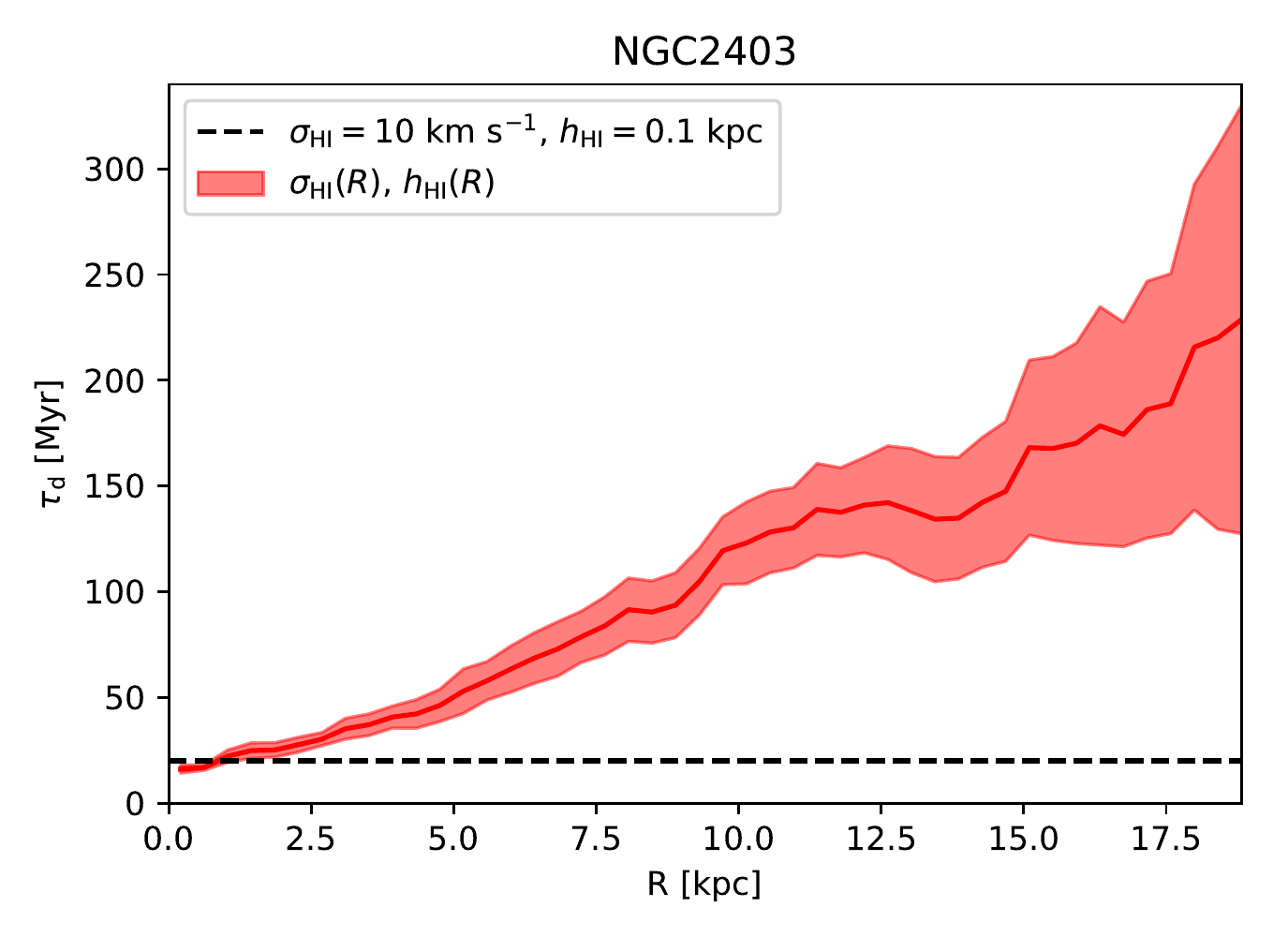}
\caption{Dissipation timescale of HI turbulence as a function of the galactocentric radius calculated with Eq.~\ref{eq:tau_d_def} for NGC~2403. 
The dashed black line is $\tau_\mathrm{d}=20$~Myr obtained with a constant velocity dispersion of 10 \kms and a constant scale height of 0.1~kpc (i.e. no flaring). 
The red curve and band are $\tau_\mathrm{d}(R)$ and its uncertainty adopting the radially-decreasing $\sigma_\mathrm{HI}(R)$ and the increasing $h_\mathrm{HI}(R)$ from 
\citetalias{2019Bacchini}.}
\label{fig:NGC2403tau}
\end{figure}

Taking NGC~2403 as an example, Figure~\ref{fig:NGC2403tau} shows the dramatic effect of including the flaring of the HI when deriving the dissipation timescale as a 
function of the galactocentric radius (Eq.~\ref{eq:tau_d_def}). 
The dashed black line is $\tau_\mathrm{d}$ obtained with $\sigma_\mathrm{HI}=10$ \kms and $h_\mathrm{HI}=100$~pc, which gives a constant dissipation timescale of 20~Myr. 
The red curve represents $\tau_\mathrm{d}(R)$ calculated with the profiles of $h_\mathrm{HI}(R)$ and of $\sigma_\mathrm{HI}(R)$ from \citetalias{2019Bacchini}. 
At large radii, the dissipation timescale is prolonged by one order of magnitude with respect to the constant $\tau_\mathrm{d}$. 
In other words, the observed turbulent energy is easier to maintain in thick discs, even with few SN explosions. 
We note that the timescale shown in Fig.~\ref{fig:NGC2403tau} should be considered a lower limit in the framework of Kolmogorov's theory, as the observed velocity dispersion used in 
this example still includes the contribution of thermal motions. 

The rate of kinetic energy injection per unit area and per unit time injected by multiple SN explosions is \citep[e.g.][]{2009Tamburro,2019Utomo}
\begin{equation}\label{eq:E_SNe_rate}
\dot{E}_\mathrm{turb,SNe} =\eta \mathrm{\mathcal{R}_\mathrm{cc}} E_\mathrm{SN} \, ,
\end{equation}
where $\eta$ is the dimensionless efficiency of SNe in transferring kinetic energy to the ISM, $\mathcal{R}_\mathrm{cc}$ is the rate of core-collapse SNe per unit area, 
and $E_\mathrm{SN}=10^{51}$~erg is the total energy released by a single SN. 
The SN rate per unit area can be obtained from the SFR surface density as 
\begin{equation}\label{eq:Rcc}
\mathcal{R}_\mathrm{cc}=\Sigma_\mathrm{SFR} f_\mathrm{cc} \, ,
\end{equation}
where $f_\mathrm{cc}$ is the number of 
core-collapse SNe that explode for unit of stellar mass formed. 
This latter is $f_\mathrm{cc} \approx 1.3 \times 10^{-2}$~\msun$^{-1}$ for a Kroupa initial mass function \citep{2002Kroupa}. 
In Sec.~\ref{sec:discussion_caveats}, we discuss the effect of including type Ia SNe and different parameters for the initial mass function. 

The turbulent energy per unit area from SN feedback is obtained from Eq.~\ref{eq:tau_d_def} using Eq.~\ref{eq:Ld_2h}, Eq.~\ref{eq:E_SNe_rate}, and Eq.~\ref{eq:Rcc}
\begin{equation}\label{eq:E_SNe}
\begin{split}
E_\mathrm{turb,SNe} & =\eta \Sigma_\mathrm{SFR} f_\mathrm{cc} E_\mathrm{SN}  \frac{2 h}{\upsilon_\mathrm{turb}} 
		     \simeq \eta \left( 2.6 \times 10^{46} \text{erg pc}^{-2} \right) \\
		    & \left( \frac{\Sigma_\mathrm{SFR}}{10^{-4} \text{\msun} \text{yr}^{-1} \text{kpc}^{-2}} \right) 
		     \left( \frac{h}{100 \text{ pc}} \right) 
		    \left( \frac{\upsilon_\mathrm{turb}}{10 \text{\kms}} \right) ^{-1} \, ,
\end{split}
\end{equation}
which gives the turbulent energy component in Eq.~\ref{eq:E_mod_def}. 
In Sec.~\ref{sec:method_HB}, we describe how we estimated $\eta$ for our galaxies using the observational constraints obtained in Sec.~\ref{sec:obs}. 

\subsubsection{Thermal energy}\label{sec:method_Eth}
The thermal velocity in Eq.~\ref{eq:sigma_obs_def} mainly depends on the temperature of the gas $T$
\begin{equation}\label{eq:vth_def}
 \upsilon_\mathrm{th} = \sqrt{ \frac{ k_\mathrm{B} T}{ \mu m_\mathrm{p}}} 
 \simeq \left(9.1 \text{\kms} \right) 
 \left( \frac{T}{10^4 \text{ K}} \right)^{\frac{1}{2}} 
 \mu^{-\frac{1}{2}} \, ,
\end{equation}
where $\mu$ is the mean particle weight in units of the proton mass $m_\mathrm{p}$ and $k_\mathrm{B}$ is the Boltzmann constant. 
The mean particle weight varies with the chemical composition of the emitting particles: it is $\mu \approx 1$ for HI, and $\mu \approx 28$ for CO. 
The thermal energy per unit area in Eq.~\ref{eq:E_mod_def} is then
\begin{equation}\label{eq:E_th_def}
\begin{split}
 E_\mathrm{th}(R) &=\frac{3}{2} \Sigma(R) \frac{ k_\mathrm{B} T}{ \mu m_\mathrm{p}} \\
		     & \simeq \left( 2.5 \times 10^{46} \text{ erg} \, \text{pc}^{-2} \right) 
		      \left( \frac{\Sigma}{10 \text{ \msun pc}^{-2}} \right)
		      \left( \frac{T}{10^4 \text{ K}} \right) 
		      \mu^{-1} ,
\end{split}
\end{equation} 
where $\Sigma(R)$ is the surface density of the gas (atomic or molecular), $\mu \approx 1.36$ for the atomic gas, and $\mu \approx 2.3$ for the molecular gas. 
Thanks to observations and physical models of the ISM, we have useful constraints on the temperature distribution of the atomic gas and the molecular gas in galaxies 
\citep*[see \S 4.2 and \S D.2.2 in][]{2019TheBook} and we can estimate the contribution of thermal motions to the observed velocity dispersion. 

\subsubsection*{Atomic gas}\label{sec:method_HI}
The atomic gas is distributed in two phases, CNM and WNM \citep[e.g.][]{1995Wolfire,2003Wolfire,2003Heiles}, whose contribution to the total thermal velocity depends on 
their abundance. 
Let us define $f_\mathrm{w}$ as the fraction of warm atomic gas and label the rest as CNM (i.e. $1-f_\mathrm{w}$). 
The thermal energy in Eq.~\ref{eq:E_mod_def} is then 
\begin{equation}\label{eq:E_th_def_2phase}
 E_\mathrm{th}=E_\mathrm{th,c} + E_\mathrm{th,w} 
 = \frac{3}{2} \Sigma_\mathrm{atom} \left[ \left(1-f_\mathrm{w} \right) \upsilon_\mathrm{th,c}^2 +  f_\mathrm{w} \upsilon_\mathrm{th,w}^2  \right] \, , 
\end{equation}
where $E_\mathrm{th,c}$ and $E_\mathrm{th,w}$ are the thermal energy of the CNM and the WNM respectively, which can be calculated using Eq.~\ref{eq:E_th_def} if their 
temperatures are known. 
In Eq.~\ref{eq:sigma_obs_def}, we have then
\begin{equation}\label{eq:vth_def_2phase}
 \upsilon_\mathrm{th}= \sqrt{ f_\mathrm{w} \upsilon_\mathrm{th,w}^2 +  \left(1-f_\mathrm{w} \right) \upsilon_\mathrm{th,c}^2 } \, ,
\end{equation}
where $\upsilon_\mathrm{th,c}$ and $\upsilon_\mathrm{th,w}$ are given by Eq.~\ref{eq:vth_def}.

We take as references the average temperatures of the atomic gas resulting from the model by \cite{2003Wolfire} (see their Table 3), which are distributed between $T\approx 40$~K and 
$T\approx$ 190~K for the CNM, and $T\approx 7000$~K and $T\approx 8830$~K for the WNM. 
These temperature ranges correspond to $0.6 \text{ \kms} \lesssim \upsilon_\mathrm{th,c} \lesssim 1.3 \text{ \kms}$ for the cold HI and 
$7.6 \text{ \kms} \lesssim \upsilon_\mathrm{th,w} \lesssim 8.6 \text{ \kms}$ for the warm HI, respectively. 
This approach implicitly assumes that the HI is thermally stable, while there are observational indications that $\approx 50$ \% of the HI in our Galaxy is in the 
thermally unstable state with $T \approx$500--5000~K \citep{2003Heiles}. 
Given these uncertainties, we decided to consider two extreme cases, the first with $f_\mathrm{w}=0$ and the second with $f_\mathrm{w}=1$, which are analised in 
Sec.~\ref{sec:results_coldHI} and Sec.~\ref{sec:results_warmHI}. 
Clearly, in these two particular cases, we can directly use Eq.~\ref{eq:vth_def} and Eq.~\ref{eq:E_th_def}. 
This choice allows us to test whether SN feedback can maintain turbulence even with the minimum possible contribution from thermal motions (i.e. CNM case, $f_\mathrm{w}=0$) and to quantify 
the dependence of our results on the HI temperature distribution. 
In Sec.~\ref{sec:results_SNetwophase}, we investigate the two-phase scenario with $0<f_\mathrm{w}<1$ and using Eq.~\ref{eq:E_th_def_2phase} and Eq.~\ref{eq:vth_def_2phase}. 

\subsubsection*{Molecular gas}\label{sec:method_H2}
Molecular gas is observed in giant molecular clouds, where the temperatures are typically very low ($T \approx 10-15$~K). 
This gas can be sightly warmer close to young stars \citep[e.g.][]{2017Redaelli}, but the coldest fraction is undoubtedly dominant in mass. 
We chose $10 \text{ K} \lesssim T \lesssim 15 \text{ K}$, which for CO roughly corresponds to $\upsilon_\mathrm{th} \approx 0.06 \pm 0.005 $ \kms. 
Hence, thermal motions do not significantly contribute to the observed velocity dispersion (i.e. 5--15 \kms) for the typical temperatures in molecular clouds. 

\subsection{Comparison of the model with the observations}\label{sec:method_HB}
In this section, we summarise the method used to infer the SN feedback efficiency required to maintain the turbulent energy in our galaxies. 
Further details on the formalism of this method can be found in Appendix~\ref{ap:HBay}. 

The algorithm works in the same way for the four cases under consideration: 
$i)$ cold atomic gas, 
$ii)$ warm atomic gas, 
$iii)$ two-phase atomic gas, and 
$iv)$ molecular gas (for 7 galaxies). 
We note that the third case involves two free parameters, namely the SN efficiency and the fraction of WNM $f_\mathrm{w}$. 
We assume that the velocity dispersion of the gas is given by the contribution of thermal and turbulent motions (i.e. Eq.~\ref{eq:sigma_obs_def}), and that turbulence is 
entirely driven by SNe, implying $0<\eta<1$ in Eq.~\ref{eq:E_SNe}. 
The efficiency is assumed to be constant with the galactocentric radius, but it is allowed to vary from one galaxy to another. 
The same is for $f_\mathrm{w}$ in the case of two-phase atomic gas. 
We adopted a hierarchical Bayesian approach \citep[e.g.][]{2019Delgado,2019Cannarozzo,2019Lamperti} to compare a model for the energy components to the observed profiles. 

Bayesian inference is based on the Bayes' theorem 
\begin{equation}\label{eq:bayes_th}
 p ( \Theta | \mathcal{D} ) \propto p( \mathcal{D} | \Theta ) p( \Theta ) \, ,
\end{equation}
where $p ( \Theta | \mathcal{D} )$ is the probability distribution of a model depending on a set of parameters $\Theta$ given the data $\mathcal{D}$, $p( \mathcal{D} | \Theta )$ is the 
probability distribution of the data given a set of parameters (i.e. the likelihood), and $p( \Theta )$ is the prior distribution of the parameters, which includes our a-priori knowledge 
about their value. 
The Bayesian approach allows us to take into account the uncertainties on the observed quantities (including the upper limits on $\Sigma_\mathrm{SFR}$), the priors, 
and the correlation between the model parameters (see for example the case of the two-phase atomic gas in Sec.~\ref{sec:results_SNetwophase}). 
Thus, we obtain a posterior distribution on $\eta$ which is marginalised over all the other parameters of the model. 

Hierarchical methods allow a further level of variability, as the priors on the model parameters depend on an additional set of parameters, the hyper-priors. 
In other words, the parameters of the priors ($\Phi$) are sampled as the other parameters of the model, assigning them hyper-prior distributions $p (\Phi)$. 
Thus, the Bayes' rule is written as 
\begin{equation}\label{eq:bayes_th_HB}
 p ( \Theta | \mathcal{D} ) \propto p( \mathcal{D} | \Theta ) p( \Theta | \Phi ) p( \Phi ) \, ,
\end{equation}
where $p( \Theta | \Phi )$ is the probability distributions of the priors given the hyper-priors. 
This allows us to parametrise the uncertainty on the priors and use it to obtain robust errors on the final value of $\eta$. 

In practice, the observed $\Sigma_\mathrm{SFR}$ and $h$\footnote{We implicitly consider the scale height $h$ as observed data, even if it is 
derived assuming the hydrostatic equilibrium \citepalias[see][]{2019Bacchini}.}
are considered as realisations of normal distributions centered on unknown true values (i.e. $\Sigma_\mathrm{SFR}^T$ and $h^T$) and with standard deviation 
given by the uncertainty on the measurements (i.e. $\Delta \Sigma_\mathrm{SFR}$ and $\Delta h$). 
The true values are assumed to have a log-normal distribution, which depends on priors and hyper-priors (see Appendix~\ref{ap:HBay}). 
For any given galaxy, $\Sigma_\mathrm{SFR}^T$ and $h^T$ are compared with the observed values $\Sigma_\mathrm{SFR}$ and $h$ in order 
to obtain the probability of the observed values given the true values (i.e. the likelihood). 
For four galaxies in the sample (i.e. DDO~154, NGC~2403, NGC~3198, and NGC~6946), the values of the observed SFR surface density at large radii are upper limits, 
hence $\Sigma_\mathrm{SFR}^T$ is assumed to be a uniform distribution from 0 to the upper limit (see Sec.~\ref{sec:obs_SFRD}). 

Similarly, the observed velocity dispersion $\sigma$ is compared to the distribution of a true velocity dispersion $\sigma^T$ to calculate the likelihood probability. 
In particular, $\sigma^T$ is obtained through Eq.~\ref{eq:sigma_obs_def}, which requires to model the thermal and the turbulent velocity components. 
We have seen in Sec.~\ref{sec:method_Eth} that the thermal equilibrium models of the ISM provide useful constraints on the temperature ranges of the gas and the contribution of 
thermal motions. 
Hence, we assume that the center and the standard deviation of the thermal velocity distribution are different for the three single-phase cases under consideration: 
$i)$ 1 \kms and 0.4 \kms for the cold atomic gas,
$ii)$ 8.1 \kms and 0.5 \kms for the warm atomic gas,  and
$iv)$ 0.06 \kms and 0.005 \kms for the molecular gas. 
For the two-phase case, the thermal velocity is modelled as a uniform distribution between 1 \kms and 8.1 \kms. 
Our a priori knowledge about the turbulent motions is instead very limited, hence we define $\upsilon_\mathrm{turb}^T$ as a log-normal distribution characterised by weakly informative 
priors and hyper-priors on its centroid velocity and standard deviation (see Appendix~\ref{ap:HBay}). 

A fourth observed quantity, namely the gas surface density $\Sigma$, is available to constrain the parameters of our model for the theoretical energy of the gas. 
We can use $\upsilon_\mathrm{turb}^T$ to calculate two useful distributions: $a)$ the turbulent energy per unit mass $3/2 (\upsilon_\mathrm{turb}^T){^2}$, and 
$b)$ the energy from SNe $E_\mathrm{turb,SNe}$ (Eq.~\ref{eq:E_SNe}). 
This latter is derived using a uniform prior for $\eta$, whose value is between 0 and 1 (see Appendix~\ref{ap:HBay}). 
By dividing $E_\mathrm{turb,SNe}$ by the density-normalised turbulent energy, we obtain a ``prediction'' for the gas surface density that can be compared with the observed one, 
which is modelled as a normal distribution centered on $\Sigma$ and with standard deviation $\Delta \Sigma$. 
From this comparison, we infer the posterior probability of the observed values given this prediction for the surface density. 

Finally, we obtain the posterior distribution of $\eta$ by marginalising over all the other parameters of the model and calculate its median value, whose uncertainty is given 
by the 16th and the 84th percentiles of the posterior distribution. 
Henceforth, we refer to this median value as the ``best efficiency''. 
In the case of the two-phase atomic gas, we derive the posterior distribution of $f_\mathrm{w}$ and its median (i.e. best) value as well. 
These best values characterise the ``best model'' in each case under consideration. 
For each best model, we extract the posterior distributions of $E_\mathrm{th,c}(R)$, $E_\mathrm{th,w}(R)$, $E_\mathrm{th}(R)$, $E_\mathrm{turb,SNe}(R)$, and $E_\mathrm{mod}(R)$ shown in 
Sec.~\ref{sec:results} (see Appendix~\ref{ap:HBay}), and of $\upsilon_\mathrm{turb}^T(R)$, which is used in Sec.~\ref{sec:results_warmHI} to analyse the Mach number in the case of warm atomic gas. 
The errors on these quantities are calculated as the 16th and the 84th percentile of the posterior distributions. 
The parameter space is explored with the Hamiltonian Monte Carlo sampler implemented in the Python routine \texttt{PyMC3} \citep{2011Hoffman,2016Salvatier}. 

\section{Results}\label{sec:results}
In this section, we present the results of our analysis and focus on the best values of the SN feedback efficiency required to sustain turbulence in the atomic gas and the 
molecular gas of our sample of galaxies. 
As already mentioned, we first explore two extreme single-phase cases for the atomic gas, one with CNM only and the other with WNM only, aiming to derive a robust range of 
values for $\eta$. 
Second, we analyse the more realistic two-phase atomic gas, in which we attempt to derive not only the SN efficiency for each galaxy but also the fraction of WNM. 
Lastly, we consider the case of the molecular gas. 
We expect to find different values of $\eta$ according to the case under consideration, hence we adopt a different nomenclature in each situation: 
$\eta_\mathrm{atom,c}$ for the CNM, $\eta_\mathrm{atom,w}$ for the WNM, $\eta_\mathrm{atom,2ph}$ for the 2-phase atomic gas, and $\eta_\mathrm{mol}$ for the molecular gas.

\subsection{Cold atomic gas}\label{sec:results_coldHI}
In the case of all atomic gas in the CNM phase (i.e. $f_\mathrm{w}=0$), thermal motions give the least possible contribution to the total energy and the turbulent energy is dominant. 
Hence, the efficiency $\eta_\mathrm{atom,c}$ can be seen as an upper limit. 

\begin{figure*}
	\includegraphics[width=2.\columnwidth]{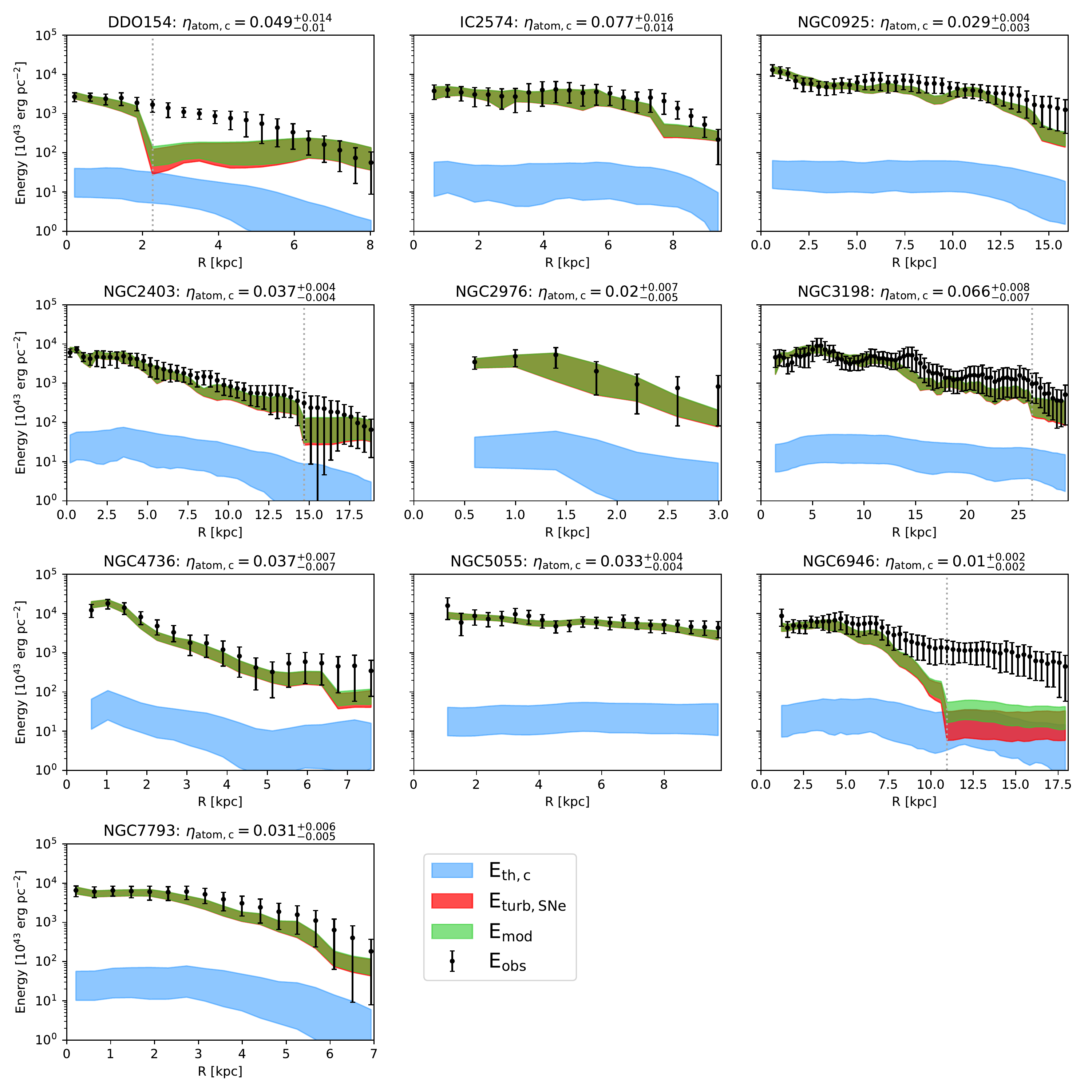}
	\caption{Observed kinetic energy per unit area of the atomic gas (black points) for our sample of galaxies 
		(the errors are calculated with the uncertainty propagation rules applied to Eq.~\ref{eq:E_obs_def}). 
		The blue area shows the thermal energy ($E_\mathrm{th,c}$) of atomic gas if it is assumed to be only CNM. 
		The red area is the turbulent energy injected by SNe ($E_\mathrm{turb,SNe}$) with the efficiency $\eta_\mathrm{atom,c}$ reported on top of each panel. 
		The green area represents the total energy ($E_\mathrm{mod}$) calculated as the sum of $E_\mathrm{th,c}$ and $E_\mathrm{turb,SNe}$ (see Sec.~\ref{sec:method_HB} and Appendix~\ref{ap:HBay}). 
		The observed profiles are well reproduced by the theoretical energy for almost all the galaxies. 
		The grey dotted vertical line, when present, indicates the outermost radius with measured $\Sigma_\mathrm{SFR}$, hence the upper limit is used for larger radii.}
	\label{fig:allgals_results_SNe_HI_cnm}
\end{figure*}

Figure~\ref{fig:allgals_results_SNe_HI_cnm} shows, for the galaxies in our sample, the observed energy $E_\mathrm{obs}$ (black points, Eq.~\ref{eq:E_obs_def}) and the total kinetic energy $E_\mathrm{mod}$ of the best model (green area, Eq.~\ref{eq:E_mod_def}) using the SFR surface density from \cite{2008Leroy} and \cite{2010Bigiel}. 
$E_\mathrm{mod}$ is the sum of the thermal energy $E_\mathrm{th,c}$ (blue area, Eq.~\ref{eq:E_th_def}) and the turbulent energy injected by SNe $E_\mathrm{turb,SNe}$ (red area, Eq.~\ref{eq:E_SNe}). 
We note that the turbulent energy is two orders of magnitude higher than thermal energy of the CNM at all radii, hence the areas representing $E_\mathrm{mod}(R)$ and $E_\mathrm{turb,SNe}(R)$ 
tend to overlap. 
The dotted grey vertical line indicates, for DDO~154, NGC~2403, NGC~3198, and NGC~7793, the outermost radius with measured $\Sigma_\mathrm{SFR}$, hence the upper limit is used for 
the radii beyond. 

The profiles of the observed energy are well reproduced by the theoretical total energy for almost all the galaxies. 
The efficiencies are $ \lesssim 0.08$ and their median value is $\langle \eta_\mathrm{atom,c} \rangle \approx 0.035$ (see Table~\ref{tab:all_eta}), showing that SNe with low efficiency 
can sustain turbulence. 
However, our model cannot fully reproduce the observed energy in the region $2 \text{ kpc}\lesssim R \lesssim 5 \text{ kpc}$ of DDO~154 and in the outskirts of IC~2574 and NGC~6946, 
indicating that some contribution from the thermal energy of the WNM may be required. 
We stress that the assumption that all the atomic gas is in the form CNM is extreme and unrealistic, so in this case we are underestimating the thermal contribution. 
Using the SFR surface density from \cite{2009MunozMateos}, we obtain efficiencies that are, on average, a factor of $\sim 2$ lower than the values found using the profiles from 
\cite{2008Leroy} and the median is indeed $\langle \eta_\mathrm{atom,c} \rangle \approx 0.015$ (see Table~\ref{tab:all_eta}). 

We note that turbulent motions are supersonic in the case of cold atomic gas (i.e. $\upsilon_\mathrm{turb} \gg \upsilon_\mathrm{th,c}$). 
Since the ISM is compressible, supersonic turbulence produces shocks that transform kinetic energy into internal energy, which may also be lost radiatively 
\citep[see e.g.][\S~11]{2005Tielens}. 
This would invalidate the energy conservation in the inertial range of the cascade assumed in our model, which is based on the implicit assumption of gas incompressibility in order to 
apply the Kolmogorov's theory. 
In the supersonic regime, our model may be suitable to describe the solenoidal motions, which are incompressible and conserve the energy \citep{2004ElmegreenScalo}. 
In particular, the solenoidal motions are expected to be dominant with respect to the compressible ones (see Sec.~\ref{sec:results_H2} for further discussion). 
We recall however that the case of cold atomic gas is unrealistic, hence we may expect that strong shocks have a more limited impact. 

\subsection{Warm atomic gas}\label{sec:results_warmHI}
We now consider the case of atomic gas in the warm phase (i.e. WNM), thus the thermal motions give the maximum possible contribution to the total energy. 
Hence, the efficiency of SN feedback $\eta_\mathrm{atom,w}$ can be considered a lower limit. 

\begin{figure*}
\includegraphics[width=2.\columnwidth]{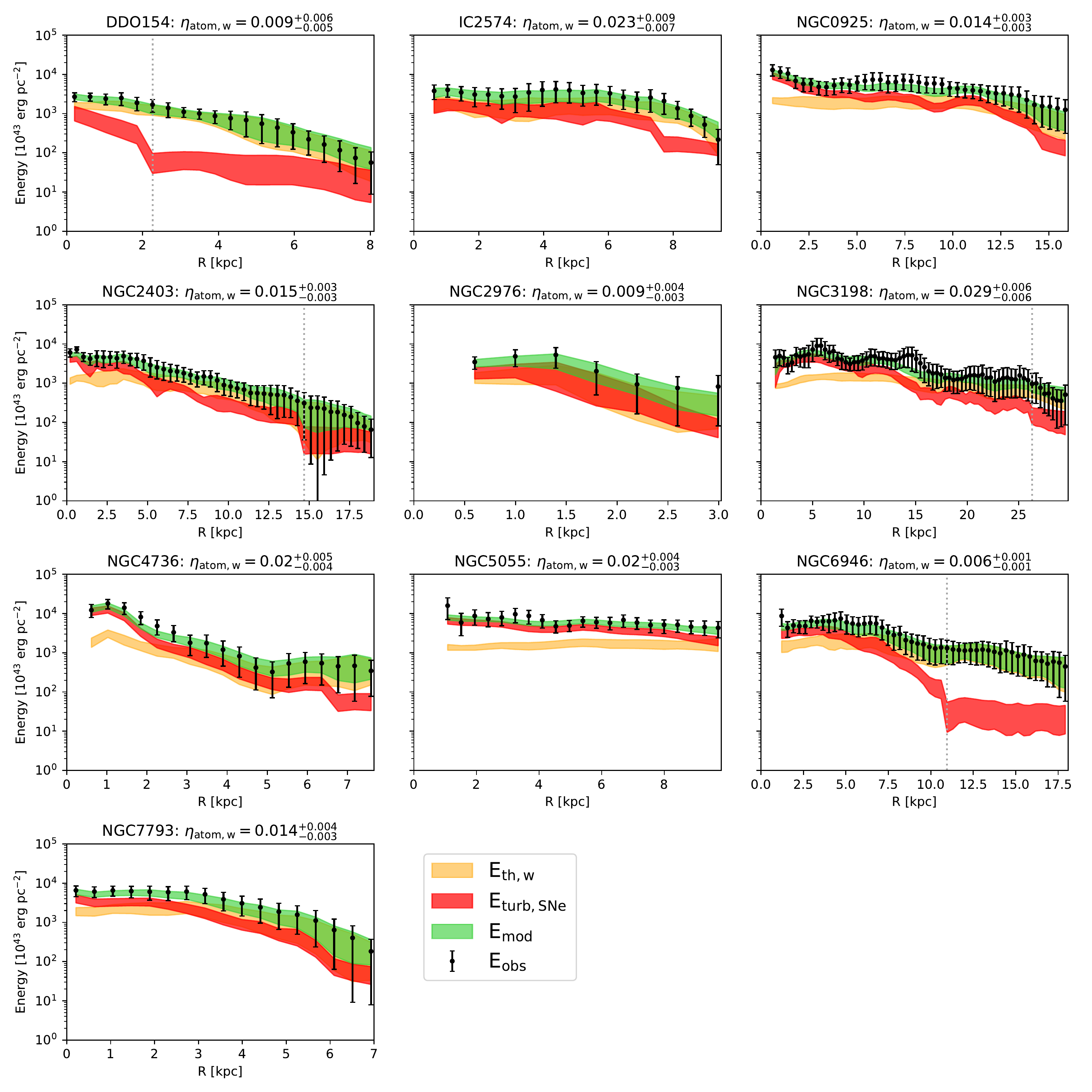}
\caption{Same as Fig.~\ref{fig:allgals_results_SNe_HI_cnm}, but the atomic gas is assumed to be only WNM in this case. The orange band shows the thermal energy ($E_\mathrm{th,w}$) and the 
best efficiency is indicated as $\eta_\mathrm{atom,w}$.}
\label{fig:allgals_results_SNe_HI_wnm}
\end{figure*}

The profiles of the observed energy are extremely well reproduced, better than in the previous case for the CNM. 
In Fig.~\ref{fig:allgals_results_SNe_HI_wnm}, we show the observed energy and the total energy of the best model using the SFR surface density from \cite{2008Leroy}. 
The main difference with respect to Fig.~\ref{fig:allgals_results_SNe_HI_cnm} is that $E_\mathrm{turb,SNe}$ dominates only in the inner regions of galaxies, while it is comparable to 
or lower than $E_\mathrm{th,w}$ (orange area) in the other parts. 
The efficiencies are generally a factor of $\sim 2$ lower than in the case of cold atomic gas and the median value is $\langle \eta_\mathrm{atom,w} \rangle \approx 0.015$ 
(see Table~\ref{tab:all_eta}). 
This indicates that SN feedback with low efficiency can maintain turbulence in the warm atomic gas and that no additional source is required. 
Using the SFR surface density from \cite{2009MunozMateos}, the resulting efficiencies are lower that those obtained with the other profiles (see Table~\ref{tab:all_eta}), 
the median is indeed $\langle \eta_\mathrm{atom,w} \rangle \approx 0.006$.

In the case of warm atomic gas, it is interesting to analyse the profile of the turbulent velocity in our galaxies to see whether turbulence is supersonic or subsonic. 
For our sample of galaxies, we calculated the Mach number ($\mathcal{M} \equiv \upsilon_\mathrm{turb} / \upsilon_\mathrm{th,w}$) as a function of the galactocentric radius using the 
posterior distribution of $\upsilon_\mathrm{turb}^T(R)$ obtained from the method described in Sec.~\ref{sec:method_HB} and the thermal velocity $\upsilon_\mathrm{th,w} \approx 8.1$ \kms. 
We found that turbulent motions are generally weakly supersonic (i.e. transonic regime): the Mach number typically reaches values of about 2--2.5 only in the innermost regions of 
spiral galaxies, while it is $ \mathcal{M} \lesssim 1$ in their outskirts and for dwarf galaxies. 
The median value is indeed $\langle \mathcal{M} \rangle \approx 1$. 
This suggests that, at least in the transonic regions, adopting Kolmogorov's theory for incompressible fluids may be acceptable in the case of warm atomic gas. 

\subsection{Two-phase atomic gas}\label{sec:results_SNetwophase}
As already mentioned, both observations and theoretical models of the ISM indicate that the atomic gas is distributed in two phases, CNM and WNM \citep[e.g.][]{2003Heiles,2003Wolfire}. 
Hence, it is interesting to investigate this scenario, as it is more realistic than the single-phase cases seen in Sec.~\ref{sec:results_coldHI} and Sec.~\ref{sec:results_warmHI}. 
We used the same method explained in Sec.~\ref{sec:method_HB}, but the thermal speed and thermal energy are calculated using Eq.~\ref{eq:E_th_def_2phase} and 
Eq.~\ref{eq:vth_def_2phase}. 
The thermal speed is the second free parameter in the model and it is used to obtain the WNM fraction $f_\mathrm{w}$ together with the SN efficiency $\eta_\mathrm{atom,2ph}$. 
This experiment has two possible outcomes: 
$i)$ if the observed velocity dispersion is lower than the thermal velocity of the WNM, the posterior distributions for $\eta_\mathrm{atom,2ph}$ and $\upsilon_\mathrm{th}$ 
(and therefore $f_\mathrm{w}$) will be well-constrained and we will calculate the median and the 1$\sigma$ uncertainty on the best parameters; 
$ii)$ if the observed velocity dispersion is higher than the WNM thermal velocity, the best model will tend to be WNM-dominated and equivalent to the case seen in 
Sec.~\ref{sec:results_warmHI}. 
NGC~2403 and NGC~4736, which we discuss below, fall into the former case, while the rest of the galaxies in the sample is compatible with having $f_\mathrm{w} \approx 1$, 
hence the resulting efficiencies are equivalent to those in Table~\ref{tab:all_eta} for the warm atomic gas. 

\begin{figure*}
	\centering
	\subfloat
	{\includegraphics[width=\columnwidth]{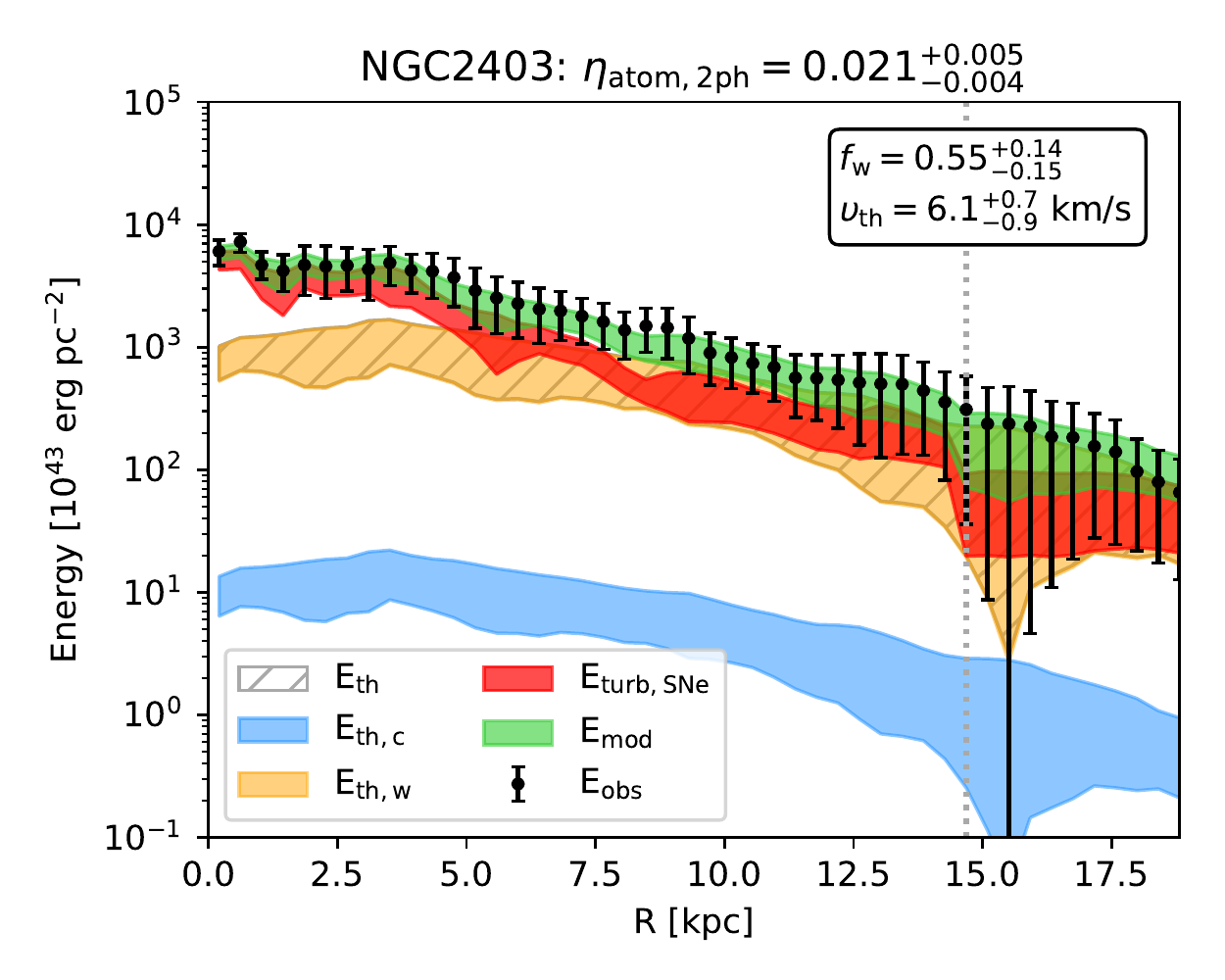}}
	\hspace{0cm}
	\subfloat
	{\includegraphics[scale=0.5]{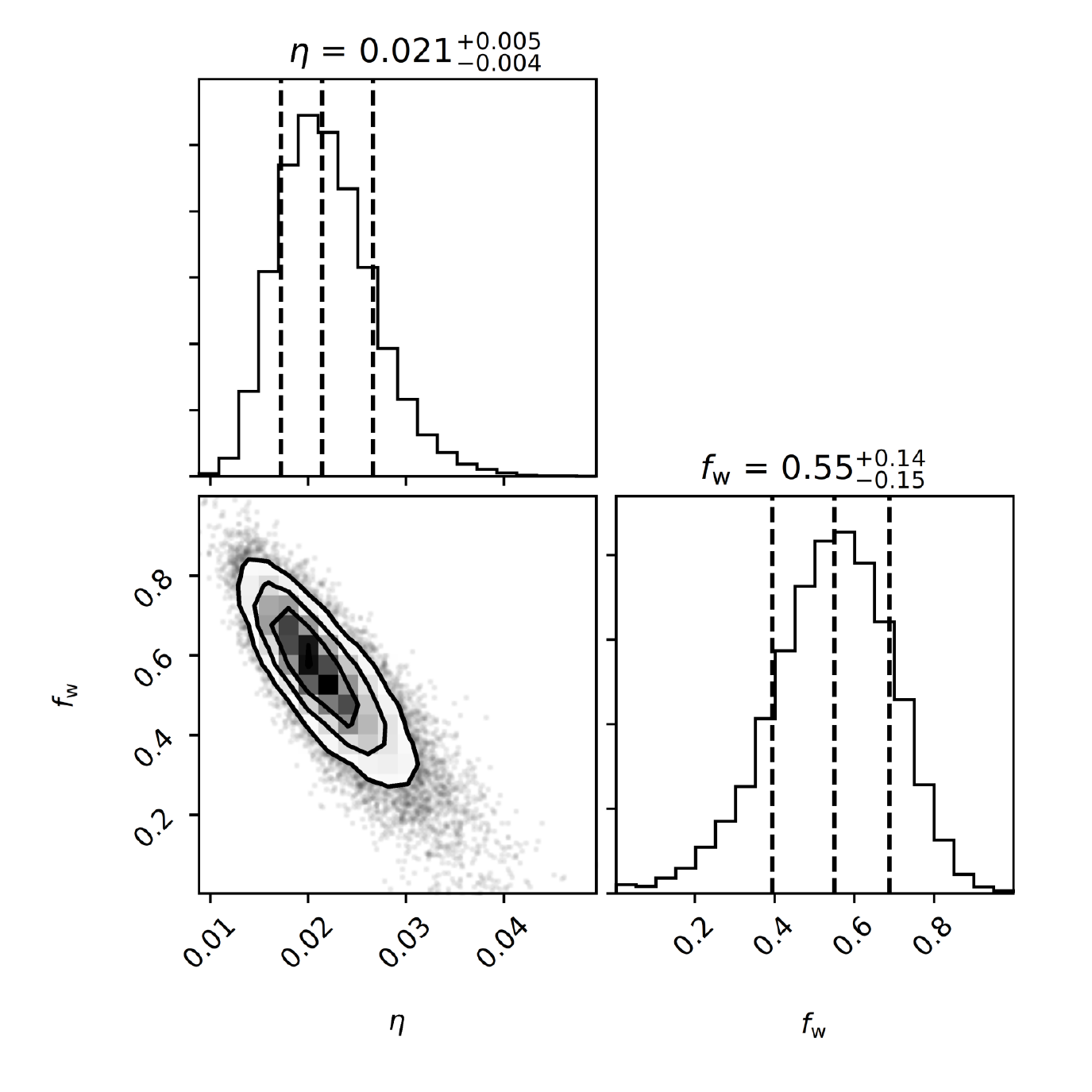}} 
	\vspace{-0.5cm}
	\subfloat
	{\includegraphics[width=\columnwidth]{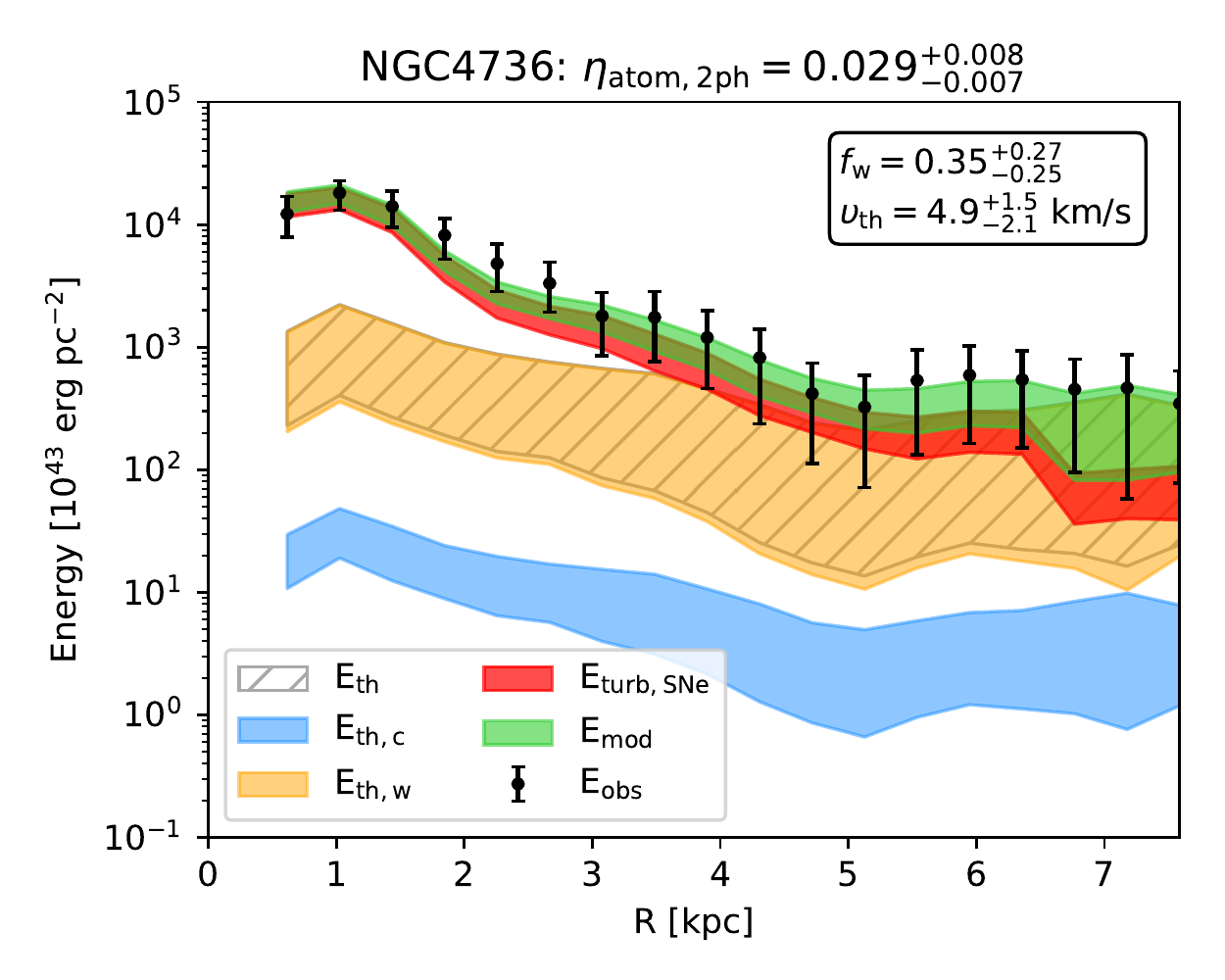}}
	\hspace{0cm}
	\subfloat
	{\includegraphics[scale=0.5]{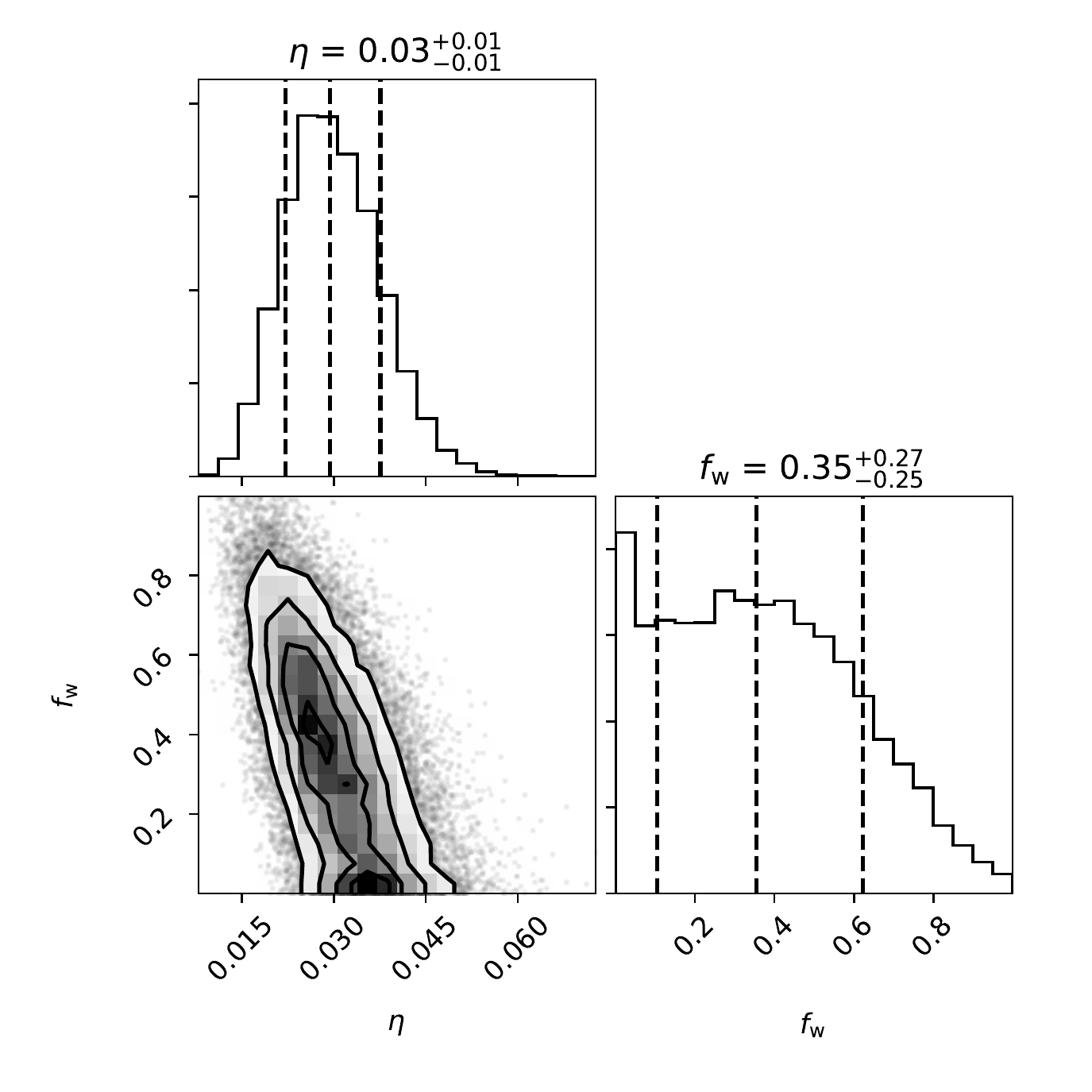}}
	\caption{\textit{Left panels:} observed kinetic energy per unit area of the atomic gas (black points; $E_\mathrm{obs}$ from Eq.~\ref{eq:E_obs_def}) for NGC~2403 (upper panel) and NGC~4736 (lower panel). The blue and the orange bands show respectively the thermal energy of the cold ($E_\mathrm{th,c}$) and the warm  ($E_\mathrm{th,w}$) atomic gas. The green area is the total kinetic energy predicted by our best model ($E_\mathrm{mod}$; see also Appendix~\ref{ap:HBay}), which includes the total thermal energy (grey dashed area; i.e. $E_\mathrm{th}$ from Eq.~\ref{eq:E_th_def_2phase}) and the turbulent energy injected by SN feedback (red area; $E_\mathrm{turb,SNe}$ from Eq.~\ref{eq:E_SNe}) with the best efficiency reported on top of the panel. The grey dotted vertical line indicates the outermost radius of NGC~2403 with measured $\Sigma_\mathrm{SFR}$, the upper limit is used for larger radii. The best fraction of WNM ($f_\mathrm{w}$) is reported in the top-right box, together with the corresponding thermal velocity of the atomic gas (Eq.~\ref{eq:vth_def_2phase}). The observed profile is very well reproduced by the theoretical energy of the best model. \textit{Right panels:} corner plot showing the marginalised posterior distributions of the SN efficiency and the fraction of WNM $f_\mathrm{w}$. The best values with 1$\sigma$ uncertainties are reported on top of each panel of the 1D posterior distribution.}
	\label{fig:2phase_panels}
\end{figure*}

The left panels in Fig.~\ref{fig:2phase_panels} show the result of this analysis for NGC~2403 and NGC~4736: the profiles of the observed energy are remarkably well reproduced by 
the two-phase best model, for both galaxies. 
We note that the SN energy is the dominant component in the inner regions, while the total thermal energy (dashed area) tends to become equally significant at large radii, as in the 
case of warm atomic gas (see Fig.~\ref{fig:allgals_results_SNe_HI_wnm}). 
The right panels in Fig.~\ref{fig:2phase_panels} show the corner plots for the $\eta_\mathrm{atom,2ph}$ and $f_\mathrm{w}$, which are obtained from the posterior distribution of 
$\upsilon_\mathrm{th}$ through Eq.~\ref{eq:vth_def_2phase}. 
From the 1D and the 2D posterior distributions, we can see that both parameters are well-constrained, despite the expected degeneracy. 
We obtain that the best model is given by $\eta_\mathrm{atom,2ph} \approx 0.021$ and $f_\mathrm{w}\approx0.55$ (or $\upsilon_\mathrm{th} \approx 6.1$ \kms) for NGC~2403, 
and $\eta_\mathrm{atom,2ph} \approx 0.029$ and $f_\mathrm{w} \approx 0.35$ (or $\upsilon_\mathrm{th} \approx 4.9$ \kms) for NGC~4736. 
It is surprising that, despite the unavoidable limitations of our approach, the estimates of the fraction of WNM are compatible with those obtained using different methods 
for the solar neighborhood \citep{2003Heiles}, the Milky Way outskirts \citep{2013Pineda}, and the Magellanic Clouds \citep{2000MarxZimmer,2000Dickey}. 
We note that the best efficiency is compatible within the uncertainties with the value obtained in the WNM-only case. 
Indeed, we expect that the thermal broadening is dominated by the WNM, even with a fraction of WNM of about 40--60\%. 
It is worth to point out that the uncertainties on $\eta_\mathrm{atom,2ph}$ should be more reliable than those obtained in the single-phase cases, as the best efficiencies is also 
marginalised on the thermal speed in this case (as this model takes into account the possible presence of CNM). 

We must keep in mind two possible caveats of this analysis. 
First, we did not include any radial or vertical gradient of the fraction of WNM, which may not be a realistic assumption for some galaxies (e.g. the Milky Way; \citealt{2013Pineda}). 
We could assume some functional form for $f_\mathrm{w}$, linear or exponential trends for instance, but this choice would introduce at least two additional free parameters 
in the model and worsen the degeneracy issue. 
The second possible caveat is the assumption that the atomic gas is thermally stable and distributed in CNM and WNM, although there are observational indications that a fraction of the 
atomic gas may be in the thermally unstable region \citep[e.g.][]{2003Heiles}. 
If we do not assume thermal equilibrium, we can use Eq.~\ref{eq:vth_def} and the best $\upsilon_\mathrm{th}$ mentioned above to estimate the temperature of the atomic gas 
distributed in a single phase. 
For NGC~2403 and NGC~4736 respectively, we obtain $T \approx 5800$~K and $T \approx 3800$~K, which both correspond to the thermally unstable regime \citep{2003Wolfire,2009Tamburro}. 

\subsection{Molecular gas}\label{sec:results_H2}
Based on the kinematic analysis of CO data cubes (see Appendix~\ref{ap:3DB}), we know that the velocity dispersion of CO is much higher than the thermal velocity expected for gas with 
$T \approx 10-15$~K (i.e. 0.05--0.07 \kms), indicating strong turbulent motions \citep[e.g.][]{2005Rosolowsky,2018Sun}. 
This fact allows us to find a very robust estimate for the SN efficiency $\eta_\mathrm{mol}$. 

\begin{figure}
	\includegraphics[width=1.\columnwidth]{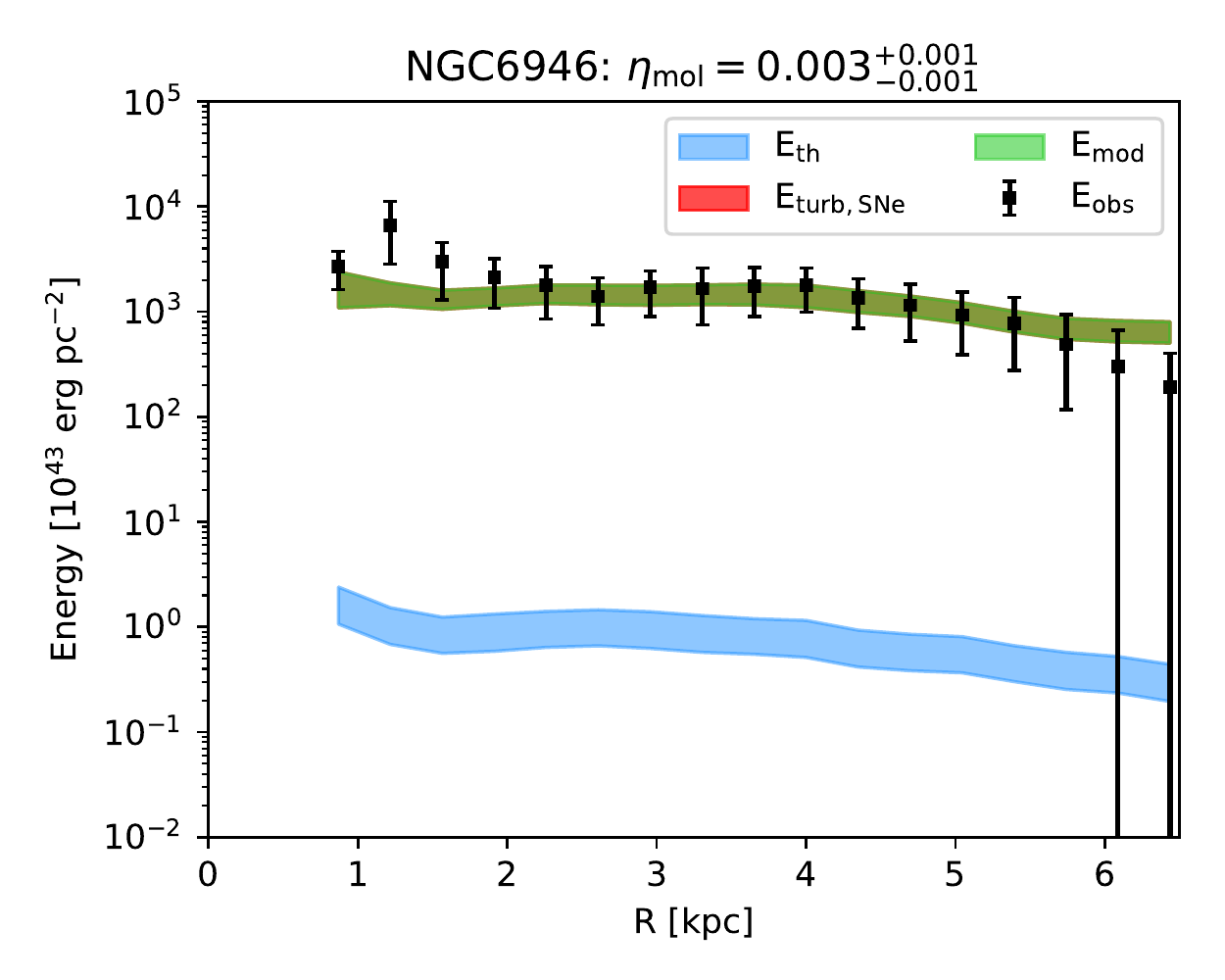}
	\caption{Observed kinetic energy per unit area of the molecular gas (black squares) for NGC~6946. 
		The green area represents the total theoretical energy ($E_\mathrm{mod}$; see also Appendix~\ref{ap:HBay}) calculated as the sum of the thermal energy (blue band; $E_\mathrm{th}$) and the turbulent 
		energy injected by SNe (red band; $E_\mathrm{turb,SNe}$) with the efficiency $\eta_\mathrm{mol}$ reported on top of the panel. 
		$E_\mathrm{turb,SNe}$ is indistinguishable from $E_\mathrm{mod}$, as the thermal energy contribution is negligible. }
	\label{fig:E_radial_H2cnm}
\end{figure}

In Fig.~\ref{fig:E_radial_H2cnm}, we show the example of NGC~6946, which has an extended molecular gas disc. 
The profile of the observed energy (black squares) is well reproduced by the theoretical profile $E_\mathrm{mod}$ with a SN efficiency of about 0.003, except for the points at 
$R\approx 1.2$~kpc and $R\approx 6.4$~kpc, which are however uncertain because of the non-circular motions in the innermost regions (see Appendix~\ref{ap:3DB}) and the low signal-to-noise 
at large radii. 
As expected, the turbulent energy is fully dominant with respect to the thermal energy (blue area). 
For the rest of our the sample, $E_\mathrm{obs}$ is also very well reproduced by models with $\eta_\mathrm{mol} \lesssim 0.016$, and the median value of the efficiency is 
$\approx 0.004$ (see Table~\ref{tab:all_eta}). 
In Table~\ref{tab:all_eta}, the shaded rows report the best efficiency values obtained using the SFR surface density from \cite{2009MunozMateos}. 
In general, the values are lower with respect to those found with the $\Sigma_\mathrm{SFR}$ from \cite{2008Leroy}, but the two estimates are compatible within the errors. 

In the case of molecular gas, turbulent motions are strongly supersonic. 
As mentioned in Sec.~\ref{sec:results_coldHI}, our model is not suitable to describe this regime, as the energy is not conserved in the turbulent cascade and Kolmogorov's theory cannot 
be applied \citep[e.g.][]{2004ElmegreenScalo}. 
Recent numerical simulations of SN-driven supersonic turbulence in molecular clouds suggest that the ratio between compressible and solenoidal motions is about $1/3-1/2$ 
(\citealt{2016Padoan,2016Pan}; see also \citealt{2017Orkisz} for an observational study). 
We could speculate that, if solenoidal and compressible motions are separable, the SN efficiency obtained with our model should by multiplied by a factor of the order of unity to take 
into account the kinetic energy dissipated through shocks. 

{\rowcolors{5}{black!15!white!50}{white}
\renewcommand{\arraystretch}{1.5}
\begin{table}
        \centering
        \caption{
        SN feedback efficiency required to sustain turbulence in the neutral gas of our sample of galaxies. 
        The values in the white rows are obtained with the SFR surface density from \cite{2008Leroy} and \cite{2010Bigiel}, while those in the shaded rows are derived with the profiles 
        from \cite{2009MunozMateos}, whose sample does not include DDO~154 (indicated with ``x''). 
        The last two rows show the median values obtain from the posteriors of all the galaxies in the sample. 
        The three columns report the best values in different cases: 
        (1) all atomic gas is CNM;
        (2) all atomic gas is WNM;
        (3) molecular gas (the ``x'' indicates that no molecular gas emission is detected).}
        \label{tab:all_eta}
        \begin{tabular}{l|c|c|c}
        \hline\hline
        Galaxy                  &\multicolumn{3}{c}{SN efficiency}	    							\\
        \hline
                                & $\eta_\mathrm{atom,c}$	& $\eta_\mathrm{atom,w}$	& $\eta_\mathrm{mol}$		\\
                                & (1)   			& (2)           		& (3)				\\
        \hline
        DDO~154                 & 0.049$^{+0.014}_{-0.010}$	& 0.009$^{+0.006}_{-0.005}$	& x				\\
				& x				& x				& x				\\
        IC~2574                 & 0.077$^{+0.016}_{-0.014}$	& 0.023$^{+0.009}_{-0.007}$	& x				\\
				& 0.021$^{+0.003}_{-0.003}$	& 0.007$^{+0.002}_{-0.002}$	& x 				\\
        NGC~0925                & 0.029$^{+0.004}_{-0.003}$	& 0.014$^{+0.003}_{-0.003}$	& 0.0004$^{+0.0002}_{-0.0001}$\\
				& 0.020$^{+0.002}_{-0.002}$	& 0.010$^{+0.002}_{-0.002}$	& 0.0004$^{+0.0002}_{-0.0001}$\\
        NGC~2403                & 0.037$^{+0.004}_{-0.004}$	& 0.015$^{+0.003}_{-0.003}$	& 0.004$^{+0.002}_{-0.002}$	\\
				& 0.013$^{+0.001}_{-0.001}$	& 0.006$^{+0.001}_{-0.001}$	& 0.002$^{+0.001}_{-0.001}$	\\
        NGC~2976                & 0.020$^{+0.007}_{-0.005}$	& 0.009$^{+0.004}_{-0.003}$	& 0.0013$^{+0.008}_{-0.007}$	\\
				& 0.012$^{+0.003}_{-0.003}$	& 0.005$^{+0.002}_{-0.002}$	& 0.0010$^{+0.0006}_{-0.0006}$	\\
        NGC~3198                & 0.066$^{+0.008}_{-0.007}$	& 0.029$^{+0.006}_{-0.006}$	& 0.016$^{+0.008}_{-0.006}$	\\
				& 0.018$^{+0.002}_{-0.002}$	& 0.010$^{+0.002}_{-0.002}$	& 0.007$^{+0.004}_{-0.003}$	\\
        NGC~4736                & 0.037$^{+0.007}_{-0.007}$	& 0.020$^{+0.005}_{-0.004}$	& 0.006$^{+0.002}_{-0.002}$	\\
				& 0.012$^{+0.002}_{-0.002}$	& 0.006$^{+0.002}_{-0.001}$	& 0.003$^{+0.001}_{-0.001}$	\\
        NGC~5055                & 0.033$^{+0.004}_{-0.004}$	& 0.020$^{+0.004}_{-0.003}$	& 0.010$^{+0.003}_{-0.002}$	\\
				& 0.027$^{+0.004}_{-0.003}$	& 0.016$^{+0.003}_{-0.003}$	& 0.009$^{+0.002}_{-0.002}$	\\
        NGC~6946                & 0.010$^{+0.002}_{-0.002}$	& 0.006$^{+0.001}_{-0.001}$	& 0.003$^{+0.001}_{-0.001}$	\\
				& 0.008$^{+0.001}_{-0.001}$	& 0.003$^{+0.001}_{-0.001}$	& 0.0024$^{+0.0005}_{-0.0004}$				\\
        NGC~7793                & 0.031$^{+0.006}_{-0.005}$	& 0.014$^{+0.004}_{-0.003}$	& x				\\
				& 0.013$^{+0.002}_{-0.002}$	& 0.006$^{+0.002}_{-0.001}$	& x				\\
        \hline
        All 			& 0.035$^{+0.029}_{-0.014}$	& 0.015$^{+0.009}_{-0.008}$	& 0.0042$^{+0.0075}_{-0.0035}$	\\
				& 0.015$^{+0.008}_{-0.005}$	& 0.006$^{+0.005}_{-0.002}$	& 0.0024$^{+0.0056}_{-0.0018}$	\\
        \hline
        \end{tabular}
\end{table}


\section{Discussion}\label{sec:discussion}
Our results show that SN feedback can maintain turbulence in the atomic gas of nearby disc galaxies with injection efficiency between 0.003 and 0.077. 
To drive molecular gas turbulence, the required efficiencies are also low ($\eta_\mathrm{mol}\lesssim 0.016$). 
Hence, turbulence can be sustained by SNe alone and no other energy sources are compulsorily required. 

\subsection{A ``global'' SN efficiency for nearby galaxies?}\label{sec:discussion_eta}
We have seen that the values of the best efficiency depend on the choice of the SFR surface density. 
In the case of the atomic gas, the efficiency depends on the assumed temperature distribution as well. 
However, finding a single value for the efficiencies may be useful, for example, to include a recipe for SN feedback  in numerical simulations and analytical models of galaxy evolution. 
We can use the posterior distributions of $\eta_\mathrm{atom,c}$ and $\eta_\mathrm{atom,w}$ to obtain this value in the case of the atomic gas (i.e. $\langle \eta_\mathrm{atom} \rangle$) 
and those of $\eta_\mathrm{mol}$ for the molecular gas (i.e. $\langle \eta_\mathrm{mol} \rangle$). 
For each galaxy, we extracted a random sub-sample of one thousand values from the posterior distributions of the efficiency in each of the cases explored in Sec.~\ref{sec:results}. 
For the atomic gas, we have four cases to consider: $i)$ CNM and $ii)$ WNM with $\Sigma_\mathrm{SFR}$ from \cite{2008Leroy} and \cite{2010Bigiel}, and 
$iii)$ CNM and $iv)$ WNM with $\Sigma_\mathrm{SFR}$ from \cite{2009MunozMateos}. 
For the molecular gas, we have only two cases, each with a different $\Sigma_\mathrm{SFR}$. 
We calculated the median and the 1$\sigma$ uncertainty using the sub-samples of the posterior distributions, finding $\langle \eta_\mathrm{atom} \rangle = 0.015_{-0.008}^{+0.018}$ for 
the atomic gas and $\langle \eta_\mathrm{mol} \rangle = 0.003_{-0.002}^{+0.006}$ for the molecular gas. 

Figure~\ref{fig:summary_HIH2_final} aims to summarise our findings. 
The left-hand side concerns the atomic gas and the right-hand side is for the molecular gas. 
The panels in the top row show the maximum SN energy (i.e. Eq.~\ref{eq:E_SNe} with $\eta=1$) on the $x-$axis and the turbulent component of the observed energy 
($E_\mathrm{obs,turb}$, i.e. Eq.~\ref{eq:E_obs_def} with the thermal energy subtracted) on the $y-$axis. 
To make all the points visible, we do not display the uncertainties and the symbols for the CNM cases are shaded (given that this scenario is also not fully realistic for the atomic gas).  
The red and the dark red lines show the relations $E_\mathrm{obs,turb} = \langle \eta_\mathrm{atom} \rangle E_\mathrm{turb,SNe}$ and 
$E_\mathrm{obs,turb} = \langle \eta_\mathrm{mol} \rangle E_\mathrm{turb,SNe}$ for the atomic gas and the molecular gas, respectively. 
The panels in the bottom row show the best efficiencies, as derived with the method described in Sec.~\ref{sec:method_HB}, in comparison with the averages $\langle \eta_\mathrm{atom} \rangle$ 
and $\langle \eta_\mathrm{mol} \rangle$. 
We can clearly see that, for the atomic gas, efficiencies above 0.1 are not required to sustain the observed turbulent energy. 
The majority of the points in the top left panel follow the relation with slope $\langle \eta_\mathrm{atom} \rangle$, indicating that an efficiency of about 0.015 may be a ``global'' value 
for the galaxies in our sample. 
Only a few points belong to the region where $\eta>0.1$, but they correspond to the CNM cases, which are not fully realistic. 
Similarly, the top right panel shows that the efficiencies for the molecular gas are lower than $\sim 0.01$ for our galaxies and that a possible ``global'' value is 
$\langle \eta_\mathrm{mol} \rangle \approx 0.003$. 
This may suggest that most of the SN energy is transferred to the atomic gas, which is typically the dominant gas phase across the galactic disc. 
\begin{figure*}
\includegraphics[width=2.\columnwidth]{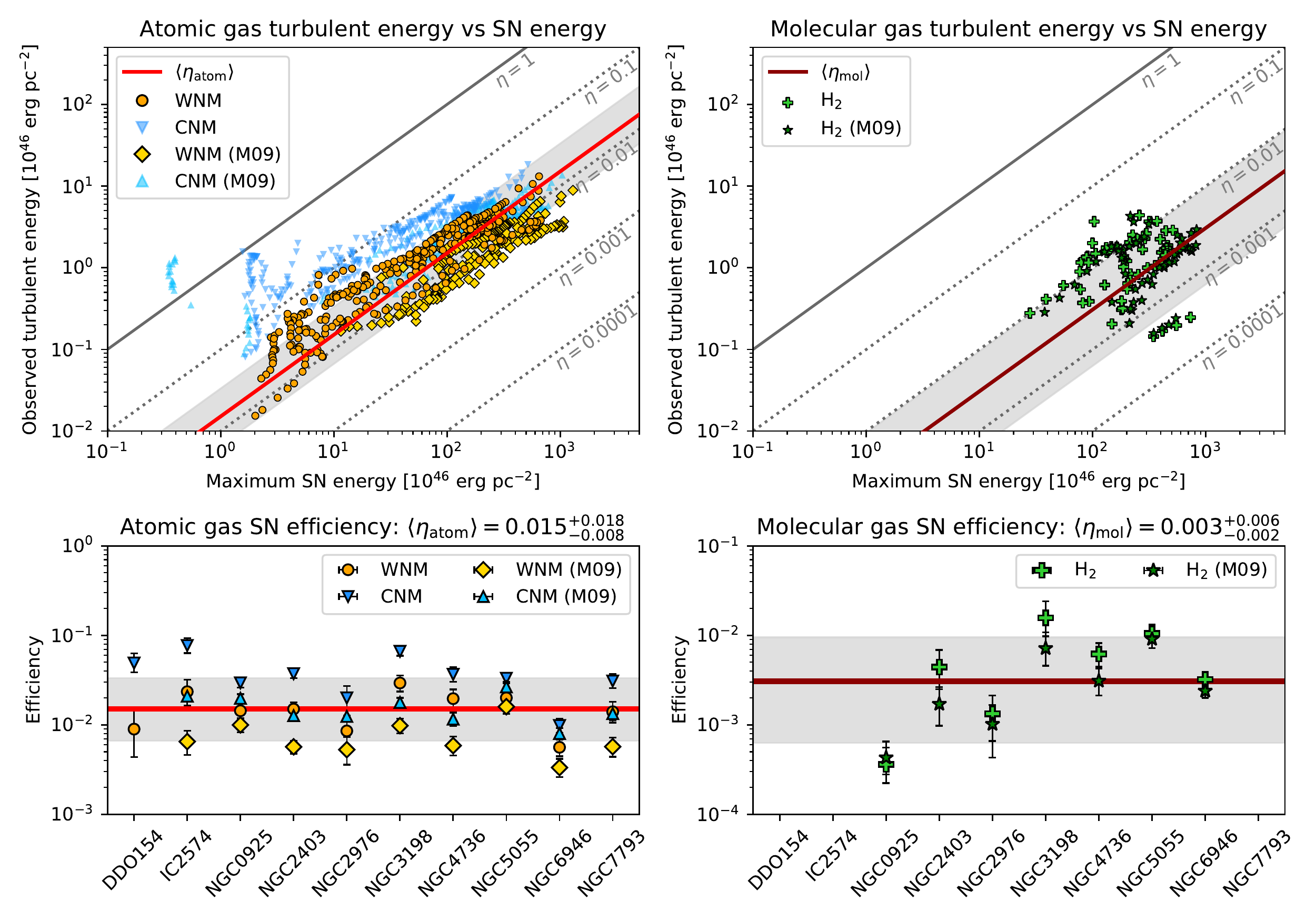}
\caption{\textit{Top row}: maximum energy provided by SNe in a dissipation timescale versus the turbulent component of the observed energy, both for the atomic gas (left) and the 
molecular gas (right). 
In the left panel, the orange points and the blue triangles are respectively for the cases of CNM and of WNM with $\Sigma_\mathrm{SFR}$ from \cite{2008Leroy} and \cite{2010Bigiel}, 
while the yellow diamonds and the light blue triangles show the corresponding cases with $\Sigma_\mathrm{SFR}$ from \cite{2009MunozMateos}. 
Each point is for a single galaxy and for a single radius. 
In the right panel, the light green crosses and the green stars are the same quantities for the molecular gas obtained with the two  $\Sigma_\mathrm{SFR}$. 
The red and the dark red lines show the relations built with the ``global'' efficiencies for the atomic gas $\langle \eta_\mathrm{atom} \rangle$ and the molecular gas 
$\langle \eta_\mathrm{mol} \rangle$ respectively (see text), with grey areas indicating the uncertainty. 
The solid grey lines show the same relation with efficiencies of 1, while the dotted lines are obtained, from top to bottom, with efficiencies of 0.1, 0.01, 0.001, and 0.0001. 
\textit{Bottom row}: summary of the best efficiency for our sample of galaxies, in the case of the atomic (left) and molecular (right) gas. 
The symbols are the same as in the top row.
The red and dark red horizontal lines are the median $\langle \eta_\mathrm{atom} \rangle$ and $\langle \eta_\mathrm{mol} \rangle$ (also indicated above each panel) with $1\sigma$ error 
(grey area). 
Overall, the results of this ``global'' analysis are consistent with those obtained with the ``spatially-resolved'' approach, showing that low-efficiency SN feedback can sustain the 
gas turbulence. }
\label{fig:summary_HIH2_final}
\end{figure*}

Theoretical and numerical models of SN explosions in the ISM tend to predict that about 10\% of the SN energy is available to feed turbulence 
\citep[e.g.][]{1974Chevalier,1998Thornton,2016Martizzi,2016Fierlinger,2019Ohlin} and some authors have found even higher values \citep[e.g. $\lesssim 25$\%;][]{2006Dib}. 
A natural question that may arise from our findings is: how is the remaining kinetic energy used ?
This residual SN energy could be spent to drive large-scale gas motions outside the disc (i.e. galactic fountain, galactic winds, outflows). 
For example, \cite{2006Fraternali,2008Fraternali} showed that the HI halo of extra-planar gas in NGC~891 and NGC~2403 could be explained with the galactic fountain cycle: a continuous 
flow of gas launched out of the disc by super-bubble blow-outs. 
They calculated that this cycle can be sustained with only a small fraction ($<4$\%) of the SN energy. 
\cite{2012Marasco} extended these studies to our Galaxy by reproducing the extra-planar HI emission with only $\approx 0.7$\% of the kinetic energy from SNe. 
Moreover, the remaining SN energy could be spent to drive galactic winds \citep[e.g.][]{2004Fraternali,2005Veilleux,2014Rubin,2017Cresci,2018DiTeodoro,2019Armillotta}. 

\subsection{Empirical evidence for the self-regulating cycle of star formation?}\label{sec:discussion_self_reg}
Taken at face value, the results presented in this work can be interpreted in a broader context, in which SN feedback and star-formation are key elements of the same self-regulating cycle 
\citep[see for example][]{1985Dopita,2011Ostriker}.
In \citetalias{2019Bacchini} and \cite{2019Bacchini_b}, we showed that the SFR volume density correlates with the total gas volume density, following a tight power-law with 
index $\approx 2$, the volumetric star formation (VSF) law, which is valid for nearby galaxies (both dwarfs and spirals) and the Milky Way. 
The observed surface densities of the gas and the SFR were converted into volume densities by dividing by the scale height derived under the assumption of hydrostatic equilibrium 
(same as here). 
The existence of the VSF law indicates that star formation is regulated by the distribution of the gas, which depends on its velocity dispersion. 
In this work, we conclude that the turbulent component of the gas velocity dispersion is driven by SN feedback. 
The energy injected into the ISM by SNe is proportional to the SFR, we can therefore imagine a cycle as follows. 
If the SFR (per unit volume) increases, the gas becomes more turbulent, implying that the gas disc thickness grows. 
The gas volume density then decreases and, according to the VSF law, the SFR consequently declines. 
This can eventually cause the support against the gravitational pull to weaken and the gas volume density to grow again, bringing to a new phase of high SFR. 
Exploring this self-regulating cycle of star formation and its role in galaxy evolution is of primary interest and we leave it to future work. 

\subsection{Comparison with previous works on SN feedback}\label{sec:discussion_previousworks}
The origin of ISM turbulence has been widely investigated in the literature (see also Sec.~\ref{sec:discussion_othersources}) using different approaches. 
In this section, we focus on two works that share some similarities with ours. 

\cite{2009Tamburro} selected a sample of 11 galaxies (5 of them are in our sample as well) and calculated the HI kinetic energy by measuring the surface density and 
the velocity dispersion using moment maps obtained from the THINGS data cubes. 
Then, they compared the observed energy with the expected turbulent energy provided by SN feedback (Eq.~\ref{eq:E_SNe}) and MRI (see discussion in Sec.~\ref{sec:discussion_MRI}). 
In particular, they assumed a constant dissipation timescale of $\tau_\mathrm{d} = 9.8$~Myr \citep{1999MacLow} for all the galaxies and constant with radius. 
They concluded that SN feedback with $\eta \lesssim 0.1$ can account for the observed kinetic energy in inner parts of the star-forming disc where 
$\Sigma_\mathrm{SFR}>10^{-3}$~M$_\odot$yr$^{-1}$kpc$^{-2}$.  
In these regions, neither MRI nor thermal motions could explain the observed velocity dispersion of atomic gas. 
At larger radii instead, they found that unphysical values for the SN efficiency $\eta \gtrsim 1$ are required to maintain the observed line broadening and kinetic energy, given the low SFR 
(i.e. $\Sigma_\mathrm{SFR}<10^{-3}$~M$_\odot$yr$^{-1}$kpc$^{-2}$) \citep[see also][]{2013Stilp}. 
Hence, \citeauthor{2009Tamburro} concluded that the HI velocity dispersion could be driven by the MRI or due to the thermal broadening associated to a warm medium with $T \approx 5000$~K.
Our results are partially in agreement with \cite{2009Tamburro} concerning the high-SFR regions of galaxies. 
However, despite we used the same data cubes as \cite{2009Tamburro}, we can reproduce the radial profiles of the observed energy per unit area with SN efficiencies $\lesssim 0.1$ 
(and no crucial help from thermal motions for most galaxies) not only in the high-SFR regions, but also in the low-SFR ones. 
The most fundamental difference with respect to this previous work is that we use the scale height of the gas disc to calculate $\tau_\mathrm{d}$, which affects the dissipation 
timescales (see Fig.~\ref{fig:NGC2403tau}). 
Moreover, an important improvement in our work is that we measured the velocity dispersion using a 3D approach, which is more robust than the 2D method based on moment maps adopted by 
\cite{2009Tamburro}. 
2D methods perform a Gaussian fit to the line profile in each pixel to measure the line broadening, but this approach can easily fail in pixels with low signal-to-noise ratio typical of 
the outskirts of galaxies. 
$^\text{3D}$\textsc{Barolo}, after dividing the galaxy in rings, simultaneously fits the rotation velocity and the azimuthally averaged velocity dispersion 
in order to minimise the residuals between the data and the model (for each ring). 
This dramatically improves the velocity dispersion measurement with respect to the pixel-by-pixel fitting of the line profile, even for data with signal-to-noise ratio of $\approx 2$ 
\citep{2015Diteodoro}.

Recently, \cite{2019Utomo} investigated the origin of turbulence in M33, considering SNe, MRI, and accretion as possible drivers. 
Using 21-cm and CO(2--1) emission-line data cubes, they studied the kinematic properties and the distribution of atomic gas and clouds of molecular gas. 
They calculated the dissipation timescale in two ways, using a constant value (i.e. $\tau_\mathrm{d}=4.3$~Myr) and the second as $\tau_\mathrm{d} = h_\mathrm{HI}/\sigma_\mathrm{HI}$, 
where $\sigma_\mathrm{HI}$ is the HI velocity dispersion and $h_\mathrm{HI}$ is the HI scale height calculated assuming the vertical hydrostatic equilibrium \citep[see][]{2010Ostriker}. 
In the former case, they found that both SN feedback and MRI with efficiencies of 1 are required to maintain turbulence up to $R \approx 8$~kpc. 
In the latter case, the observed turbulence could instead be sustained by SNe and MRI with the efficiency of about 0.1 in the inner regions and 0.6--0.8 beyond $R=6$~kpc. 
Concerning molecular clouds, these authors obtained that the observed turbulent energy could be maintained by SN feedback with $0.001 < \eta < 0.1$. 
In agreement with \cite{2019Utomo}, our results show the importance of calculating the timescale of turbulence dissipation taking into account the increase of the scale height 
and the radial decrease of the velocity dispersion. 
The main discrepancy between this previous work and ours is that we conclude that SN feedback alone can maintain turbulence of the atomic gas with efficiencies $\lesssim 0.1$ 
(see Fig.~\ref{fig:summary_HIH2_final}). 
There are several differences between this work and ours that may all jointly explain this discrepancy. 
We discuss below only the two issues with a primary impact on the SN efficiency, as we expect the others to give a secondary contribution (e.g. method to measure 
$\sigma_\mathrm{HI}$, components of the mass model for M33, thermal energy subtraction). 
First, \cite{2019Utomo} assumed that the energy injected in the ISM by a single SN is $\approx 3.6 \times 10^{50}$ erg based on the prescription for the momentum injection 
by a SN explosion given by \cite{2015Kim}, who perfomed numerical simulations of SN explosions in a two-phase medium. 
Hence, their efficiencies should be multiplied by 0.36 to be compared with ours, reducing the discrepancy. 
Second, these authors assumed $L_\mathrm{D} = h_\mathrm{HI}$ instead of $L_\mathrm{D} = 2h_\mathrm{HI}$, which clearly contributes for another factor of 2 in the efficiency. 

Overall, the main improvement in our work with respect to the literature is that we can explain the observed turbulence with SN feedback only and with a constant efficiency across the 
galactic discs. 
The primary reason for our success is that we include the scale height in the calculation of the dissipation timescale. 
Hence, in contrast with previous authors, we find no indication that other mechanisms are compulsorily required (see Sec.~\ref{sec:discussion_othersources} for further discussions). 

\subsection{Possible caveats on the analysis and stability of the results}\label{sec:discussion_caveats}
A possible caveat on this work may be that we considered only the neutral gas components of the ISM, while the ionised gas is as well highly turbulent \citep[e.g.][]{2019Poggianti,2019Melnick}. 
Within the disc, the ionised gas is typically subdominant in mass with respect to the neutral gas, thus we do not expect that including it would significantly change our conclusions. 
Some authors investigated the possible sources of the turbulent energy in the ionised gas using the velocity dispersion of H$\alpha$ emission lines \citep[e.g.][]{2009Lehnert,2017Zhou,2019Yu,2020Varidel}, 
but it remains unclear whether SN feedback models can reproduce these observations. 
Taking into account the gas disc flaring probably helps to solve this conundrum, but it is beyond the scope of this work. 

The assumption $L_\mathrm{D} = 2 h$ (Eq.~\ref{eq:Ld_2h}) might be questionable. 
We note however that even adopting $L_\mathrm{D} = h$, for instance, our conclusions would not change as the efficiencies would be increased by a factor of 2 only, still being $\lesssim 0.1$. 
Our choice is supported by observational as well as theoretical arguments. 
Analytical and numerical models of the evolution of a SN remnant predict that the shell radius reaches about 100~pc for the typical conditions of the ISM, 
namely $n \approx 1-0.1$ cm$^{-3}$ and $\sigma \approx 6-10$ \kms (\citealt{1972Cox,1974Chevalier,1974Chevalier_b,1988Cioffi,2015Martizzi}, and \S 8.7 in \citealt*{2019TheBook}). 
However, massive stars are typically found in associations and evolve simultaneously in a small region. 
These stars produce powerful winds that sweep the ISM from the surroundings, facilitating the expansion of SN shells and generating a super-bubble, that can easily reach the size of 
the disc thickness and even blow out \citep{1989MacLow}. 
Observations of HI holes with diameter of $\approx 1$~kpc in nearby galaxies corroborates this scenario \citep[e.g.][]{1991Kamphuis,1992Puche,2008Boomsma}. 
Given that $h_\mathrm{HI} \approx 300-500$~pc \citepalias[e.g.][]{2019Bacchini}, our choice of $L_\mathrm{D} = 2 h_\mathrm{HI}$ is perfectly reasonable.  
Moreover, in the Small Magellanic Cloud, the velocity power spectrum of atomic gas suggests that $L_\mathrm{D} \approx 2.3 $~kpc \citep{2015Chepurnov}, which is consistent with our 
assumption $L_\mathrm{D} \approx 2 h_\mathrm{HI}$ if we adopt $h_\mathrm{HI} \sim 1$~kpc, as indicated by \cite{2019DiTeodoro}. 
The comparison of the observed HI morphology to simulations of dwarf galaxies seems to suggest even higher values ($L_\mathrm{D} \sim 6$~kpc, see \citealt{2005Dib}). 

We derived the turbulent energy provided by SNe adopting the prescription for the energy injection rate $\dot{E}_\mathrm{turb,SNe}$ (Eq.~\ref{eq:E_SNe_rate}) 
given by \cite{2009Tamburro}, which depends on the SN rate (Eq.~\ref{eq:Rcc}). 
This latter takes into account only core-collapse SNe, as they can be directly related to the recent star formation traced by FUV emission from massive stars younger than 100~Myr 
\citep[e.g.][]{2012KennicuttEvans}. 
The fraction of SNe Ia is expected to be less than or equal to the fraction of core-collapse SNe depending on the galaxy morphological type \citep{2005Mannucci,2011Li}. 
Therefore, considering also SN Ia would not significantly change our results, but just decrease the best values of the efficiencies by a factor $\lesssim 2$, strengthening our conclusions. 
The fraction of core-collapse SNe in Eq.~\ref{eq:Rcc} also depends on the index and the upper limit on the stellar mass of the initial mass function. 
We adopted an index of $-1.3$ for the stars in the mass range between $0.1 \text{ M}_\odot$ and $0.5 \text{ M}_\odot$, and of $-2.3$ for those with mass up to $120 \text{ M}_\odot$ 
\citep{2002Kroupa}, which gives $f_\mathrm{cc} \approx 1.3 \times 10^{-2}$ \msun$^{-1}$. 
By decreasing the index or the upper limit on the stellar mass, we would obtain less massive stars and a lower $f_\mathrm{cc}$. 
However, the effect of these variations on the SN efficiency is not straightforward, also the conversion of far-ultraviolet and infrared emission to SFR is affected, but in the opposite 
direction (i.e. higher SFR for decreasing number of massive stars; see e.g. \citealt{2009Tamburro}), suggesting that our results are weakly influenced by the initial mass function parameters. 

We verified that the general conclusions of this work do not depend on the method used to obtain the best efficiencies that reproduce the observed kinetic energy.  In particular, we performed the analysis adopting two additional approaches. The first, which was used to carry out preliminary tests, avoids any fitting procedure and does not involve any assumption on the efficiency. For each galaxy, we simply subtracted, at each galactocentric radius, the expected thermal velocity from the observed velocity dispersion (Eq.~\ref{eq:sigma_obs_def}) in order to disentangle the turbulent velocity $\upsilon_\mathrm{turb}(R)$. This latter was used to estimate the turbulent energy component, which was then divided by the energy produced by SNe in one turbulent crossing time (i.e. Eq.~\ref{eq:E_SNe} with $\eta=1$). Thus, we obtained the radial profile of the SN efficiency (i.e. $\eta(R)$) without assuming $0<\eta<1$, hence the cases with $\eta>1$ were possible. For most of the galaxies, $\eta(R)$ had large uncertainties ($\gtrsim 50$\%), in particular at large radii, as the uncertainties on the observable quantities involved in this calculation (e.g. $\Sigma_\mathrm{SFR}$) are larger at large radii than in the inner regions of galaxies \citep[see also][]{2019Utomo}. This indicates that, using this approach, it is not possible to obtain fully satisfactory constraints on the efficiency in the outskirts of galaxies. \footnote{In Fig.~\ref{fig:allgals_results_SNe_HI_wnm}, the black points ($E_\mathrm{obs}$) are sometimes systematically above the green band ($E_\mathrm{mod}$) at large radii. This suggests that the best value of the efficiency is slightly more constrained by the inner points, which are less uncertain (i.e. have narrower priors), than those at large radii. However, this is a very minor effect and the model perfectly reproduces $E_\mathrm{obs}$ within the uncertainties; even forcing the outer parts to have more weight than the inner ones, the best efficiency would never increase by more than a factor $\lesssim 2$. Thus, our results strongly point to an efficiency nearly constant with radius.}
For each galaxy, we used $\eta(R)$ to calculate the median and the 1$\sigma$ uncertainty and found that these are, albeit very uncertain, grossly compatible with those obtained with the hierarchical method and a constant efficiency. The median values for the whole sample of galaxies are (using $\Sigma_\mathrm{SFR}$ from \citealt{2008Leroy} and \citealt{2010Bigiel}) $\langle \eta_\mathrm{atom,c} \rangle = 0.053_{-0.027}^{+0.060}$ for the cold atomic gas, $\langle \eta_\mathrm{atom,w} \rangle = 0.025_{-0.012}^{+0.030}$ for the warm atomic gas, and $\langle \eta_\mathrm{mol} \rangle = 0.004_{-0.002}^{+0.011}$ for the molecular gas. 
The second approach that we explored is based on the (non-hierarchical) Bayesian framework and consists in fitting Eq.~\ref{eq:E_mod_def} to the observed kinetic energy through the algorithm implemented in the Python module \texttt{emcee} \citep{2013ForemanMackey}. 
We took an efficiency constant with radius as a free parameter with a uniform prior between 0 and 1. 
The resulting best-fit efficiencies are generally compatible within the errors with those obtained with the hierarchical method. 
In particular, the median values for the whole sample of galaxies are (using $\Sigma_\mathrm{SFR}$ from \citealt{2008Leroy} and \citealt{2010Bigiel}) 
$\langle \eta_\mathrm{atom,c} \rangle = 0.042_{-0.012}^{+0.024}$ for the cold atomic gas, 
$\langle \eta_\mathrm{atom,w} \rangle = 0.024_{-0.009}^{+0.009}$ for the warm atomic gas, 
and $\langle \eta_\mathrm{mol} \rangle = 0.007_{-0.005}^{+0.004}$ for the molecular gas. 
We conclude that our results do not depend on the adopted statistical method. The fiducial approach described in Sec.~\ref{sec:method_HB} is preferable with respect to others, as it offers a rigorous treatment of the uncertainties.

\subsection{Other turbulence sources}\label{sec:discussion_othersources}
In this study, we found that SNe are sufficient to maintain the observed turbulence, hence we did not explore in detail other possible source of energy. 
Moreover, as we motivate in this section, the contribution from the other drivers is likely of secondary importance and more uncertain than SN feedback. 

\subsubsection{Other forms of stellar feedback}
SNe are not the only form of stellar feedback that can transfer kinetic energy to the ISM \citep[see][and references therein]{2004MacLowKlessen,2004ElmegreenScalo}. 
Proto-stellar outflows (i.e. jets and winds) can be quite powerful, but they inject energy on scales equal to or smaller than that of molecular cloud complexes. 
Hence, it seems unlikely that they could feed turbulence on the scale of galactic discs. 

O--B and Wolf-Rayet stars also produce strong winds, but only those with the highest masses carry a significant amount of kinetic energy into the ISM. 
For class O and Wolf-Rayet stars (lifetime $\sim$ 4~Myr), the most extreme winds have outflow rates $\dot{M}_\mathrm{wind} \sim 10^{-4} \, \mathrm{M}_\odot \mathrm{yr}^{-1}$ and velocities 
$V_\mathrm{wind} \approx 3000$ \kms \citep[e.g.][]{1996Puls,2000Nugis,2017Gatto}. 
Winds from less massive stars are typically characterised by $\dot{M}_\mathrm{wind} \sim 10^{-6} \, \mathrm{M}_\odot \mathrm{yr}^{-1}$ and $V_\mathrm{wind} \approx 2000$ \kms
\citep{1996Puls,2000Nugis}, hence their contribution to the ISM turbulence is likely lower, despite the longer lifetimes and higher number of these stars. 
\cite{2017Gatto} used 3D hydro-dynamic simulations to study the influence of winds and SNe from massive stars on the ISM. 
They compared the cumulative energy of winds and SN explosions and found that, in the whole wind phase, the most massive stars ($\approx 85 \, \mathrm{M}_\odot$) produce as much energy as 
or more energy than in the SN phase. 
On the other hand, less massive stars (9--20~$\mathrm{M}_\odot$) release in the wind phase about $10^2$--$10^4$ times less energy than in the SN phase. 
These less massive stars are much more numerous than the massive ones, thus SN explosions likely dominate over stellar winds after the first few Myr of the stellar population lifetime 
\citep{2004MacLowKlessen}. 

A further stellar source of energy is the ionising radiation from massive stars. 
Most of this energy ionises the diffuse medium around the stars, shaping HII regions and heating the surrounding gas. 
Ionised gas cools radiatively by emitting non-ionising photons and it contracts due to the thermal instability \citep*[see e.g.][\S 8.1.4]{2019TheBook}, possibly driving 
turbulent motions. 
For example, \cite{2002Kritsuk_a,2002Kritsuk_b} estimated that $\lesssim 7$\% of the thermal energy may be converted into kinetic energy through this mechanism.
However, as shown by \cite{2004MacLowKlessen}, the kinetic energy injected into the ISM by ionising radiation is about 2--3 order of magnitude lower than produced by SN explosions. 

Besides that, ionising radiation can transfer kinetic energy to the ISM also through the expansion of HII regions \citep[e.g.][]{2020Menon}. 
\cite{2012Walch} used 3D SPH simulations to study the effect of the ionising radiation from a single O7 star on a surrounding molecular cloud. 
They found that $\lesssim 0.1$\% of the ionising energy is converted into kinetic energy and that this form of stellar feedback can sustain turbulent motions of about 2--4 \kms 
\citep[see also][]{2006Mellema}. 
Overall, these results suggest that, if compared to SNe, the turbulent energy from ionising radiation is of secondary importance. 

\subsubsection{Magneto-rotational instability and shear}\label{sec:discussion_MRI}
Several authors have proposed that the MRI \citep{velikhov1959stability,1960Chandrasekhar,1991Balbus} may be the main source of turbulent energy in the outskirts of galaxies, as it 
generates Maxwell stresses that transfer kinetic energy from shear to the ISM turbulence \citep[e.g.][]{1995Hawley,1999Sellwood,2007Piontek}. 
The energy per unit area provided by MRI is \citep[e.g.][]{2004MacLowKlessen,2009Tamburro,2019Utomo}
\begin{equation}\label{eq:E_MRI}
\begin{split}
 E_\mathrm{turb,MRI} & \simeq \left( 5 \times 10^{43} \text{ erg pc}^{-2} \right)
			\eta_\mathrm{MRI}
			\left( \frac{h}{100 \text{ pc}} \right)^2 \\
		&	\left( \frac{\upsilon_\mathrm{turb}}{10 \text{ \kms}} \right)^{-1}
			\left( \frac{B}{6 \, \mu\text{G}} \right)^2
			\left( \frac{S}{\text{Gyr}^{-1}} \right) \, ,
\end{split}
\end{equation}
where $B$ is the magnetic field intensity and $S \equiv \left| \frac{d\Omega}{d \ln R} \right| = \left| \frac{d V_\mathrm{rot}}{d R} - \frac{V_\mathrm{rot}}{R}\right|$ is the shear rate 
in Gyr$^{-1}$, which depends on the angular frequency $\Omega \equiv V_\mathrm{rot}/R$ given the rotational velocity of the galaxy $V_\mathrm{rot}$. 
The energy provided by the MRI can become significant at large radii, hence it has been advocated to explain turbulence in the outskirt of galaxies, in addition to SNe 
\citep[e.g.][]{1999Sellwood,2009Tamburro,2019Utomo}. 
Let us take for example NGC~6946, which requires a significant contribution from the thermal energy of the warm atomic gas to explain the observed energy per unit area for 
$R \gtrsim 10$~kpc (see Fig.~\ref{fig:allgals_results_SNe_HI_wnm}). 
The ordered magnetic field is $B \approx 5-10 \, \mu \text{G}$ \citep{2007Beck}, the HI scale heigth grows from $h_\mathrm{HI} \approx 150$~pc at $R \approx 10$~kpc to 
$h_\mathrm{HI} \approx 180$~pc at $R \approx 17$~kpc, the rotation curve is approximately constant at about 200 \kms for $R \gtrsim 10$~kpc, and the turbulent velocity in these regions 
is also constant at about $ 5-7$ \kms (see \citetalias{2019Bacchini} and \citealt{2008Boomsma}). 
Using these values in Eq.~\ref{eq:E_MRI} and assuming $\eta_\mathrm{MRI}=1$, we obtain that the MRI can provide $E_\mathrm{turb,MRI} \approx 7-12 \times 10^{45} \text{ erg pc}^{-2}$ 
in the regions beyond $R \sim 10$~kpc. 
This estimate of $E_\mathrm{turb,MRI}$ is compatible with the observed energy of NGC~6946 (see Fig.~\ref{fig:allgals_results_SNe_HI_wnm}), but we required that 100\% of the MRI energy is 
transferred to the ISM. 
For example, non-ideal MHD effects (i.e. Ohmic diffusion, ambipolar diffusion, and Hall effect) can suppress the MRI instability \citep[e.g.][]{1999Warle,2004Kunz,2010Korpi,2019Riols}. 
In addition, there are indications that SN driven-turbulence can counteract the MRI in the star-forming regions \citep[e.g.][]{2013Gressel}. 
We also note that $E_\mathrm{turb,MRI}$ calculated with Eq.~\ref{eq:E_MRI} can be very uncertain. 
Indeed, the magnetic field intensity is difficult to measure precisely and there are indications that it varies between different regions of a galaxy 
\citep[e.g.][]{1996Beck,2007Beck,2008Chyzy}. 

More in general, shear from galactic rotation can transfer kinetic energy to the ISM. 
It is however not straightforwardly understood how to couple the large scales of galactic rotation to smaller scales and whether the energy input from shear is significant if compared to SN 
feedback (see e.g. \citealt{2004MacLowKlessen} and references therein). 
We conclude that the role of MRI and shear in sustaining the ISM turbulence is still unclear and, given our success in reproducing the observed energy with 
SNe only, we do not find evidence for the need of these contributions.

\subsubsection{Gravitational energy}
It has been proposed that turbulence in star-forming galaxies may be driven by gravity through gas accretion \citep[e.g.][]{2010Klessen,2010Elmegreen}. 
\cite{2010Klessen} investigated this mechanism for a sample including the MW and 11 nearby galaxies from the THINGS sample (IC~2547, NGC~4736, NGC~6946, and NGC~7793 are also in our sample). 
They estimated that the energy input rate from the accreted material (over the whole galaxy) is
\begin{equation}\label{eq:Einfall_rate}
 \dot{E}_\mathrm{turb,infall} \simeq \left( 1.3 \times 10^{40} \text{ erg s}^{-1} \right)
				  \left( \frac{\dot{M}_\mathrm{infall}}{1 \text{ M}_\odot \text{ yr}^{-1}} \right)
				  \left( \frac{V_\mathrm{infall}}{200 \text{ \kms}} \right)^2 \, ,
\end{equation}
where $\dot{M}_\mathrm{infall}$ and $V_\mathrm{infall}$ are the mass inflow rate and the infall speed. 
For each galaxy, they assumed that $\dot{M}_\mathrm{infall}$ is equal to the observed SFR of the galaxy, based on the idea that gas accretion should sustain star formation. 
For $V_\mathrm{infall}$, they used the rotation velocity as an approximation of the impact velocity of the accretion gas onto the galactic disc. 
\cite{2010Klessen} calculated the dissipation timescale of turbulence assuming $L_\mathrm{D} = 2 h_\mathrm{HI}$ and with a equation analogous to Eq.~\ref{eq:tau_d_def}. 
They found that less than 10\% of the kinetic energy from accretion is required to sustain turbulence in spiral galaxies (roughly similar with the values in Table~\ref{tab:all_eta}). 
On the other hand, they concluded that other energy sources should dominate in dwarf galaxies (e.g. IC~2547), as the expected accretion rate was too low to explain the observed energy. 

It is interesting to compare the energy per unit area provided by SN feedback (Eq.~\ref{eq:E_SNe}) and by accretion ($E_\mathrm{turb,infall}$). 
This latter can be obtained from Eq.~\ref{eq:Einfall_rate} by replacing $\dot{M}_\mathrm{infall}$ with $\dot{\Sigma}_\mathrm{infall}$, and multiplying by the dissipation timescale 
(Eq.~\ref{eq:tau_d_def}) and the infall efficiency $\eta_\mathrm{infall}$ (i.e. the fraction of the infall energy that goes into feeding turbulence). 
Assuming $L_\mathrm{D} = 2 h_\mathrm{HI}$, the ratio between the SNe and the infall energies is
\begin{equation}\label{eq:Einf_Esne_ratio}
\begin{split}
 \frac{E_\mathrm{turb,infall}}{E_\mathrm{turb,SNe}} & \simeq \left( 3.1 \times 10^{-2} \right)
							\frac{\eta_\mathrm{infall}}{\eta_\mathrm{SNe}}
							\left( \frac{V_\mathrm{infall}}{200 \text{ \kms}} \right)^2 \\
							& \left( \frac{\dot{\Sigma}_\mathrm{infall}}{10^{-4} \text{\msun} \text{yr}^{-1} \text{kpc}^{-2}} \right) 
							\left( \frac{\Sigma_\mathrm{SFR}}{10^{-4} \text{\msun} \text{yr}^{-1} \text{kpc}^{-2}} \right)^{-1} \, ,
\end{split}
\end{equation}
where the SN efficiency is defined as $\eta_\mathrm{SNe}$ to distinguish it from the infall efficiency $\eta_\mathrm{infall}$. 
If we take $\dot{\Sigma}_\mathrm{infall} = \Sigma_\mathrm{SFR}$ as suggested by \cite{2010Klessen}, the infall energy is about 2 orders of magnitude lower than the SN energy when 
$\eta_\mathrm{infall} \approx \eta_\mathrm{SNe}$. 
The velocity term in Eq.~\ref{eq:Einf_Esne_ratio} is unlikely to go in the direction of increasing $ E_\mathrm{turb,infall}/E_\mathrm{turb,SNe}$. 
For example, high-velocity clouds, which are among the candidates for accreting gas on to galaxies \citep[e.g.][]{2012Putman}, have velocities between $\approx 50$ \kms and 
$\approx 150$ \kms \citep[e.g.][]{2007Boomsma,2013Marasco}. 
The galactic fountain cycle is another possible channel for gas accretion, as fountain clouds can trigger gas condensation from the hot corona and fall back onto the disc, 
bringing new material\footnote{Fountain clouds return into the disc at approximately the same radius where their were launched in orbit, but with an angular momentum mismatch 
with respect to the gas in the disc, as they accreted low-angular momentum coronal gas. 
Therefore, this material is expected to move radially towards the inner and lower-angular momentum regions of the galaxy. 
In principle, these radial flows could contribute to feeding turbulence. 
However, \cite{2016Pezzulli} have shown that the radial velocity of this gas is of the order of 1 \kms, which is negligible with respect to the observed velocity dispersion. 
Thus, we do not expect radial motions to significantly contribute to turbulence. } (see \citealt{2017Fraternali} and references therein). 
However, fountain clouds have velocities below 100 \kms \citep[e.g.][]{2006Fraternali,2008Fraternali,2012Marasco,2019Marasco_b}, hence it is very
unlikely that they could transfer a significant amount of energy to the ISM. 
We also note that $\dot{\Sigma}_\mathrm{infall} / \Sigma_\mathrm{SFR}$ might vary with the galactocentric radius of galaxies. 
\cite{2012Marasco} found that the peak in the accretion rate of the galactic fountain lies well beyond the peak of the SFR in the Milky Way. 
\cite{2016Pezzulli} showed that, in general, this mismatch between the radial profiles of $\dot{\Sigma}_\mathrm{infall}$ and $\Sigma_\mathrm{SFR}$ is related to the deficit of 
angular momentum of the accreted gas with respect to the gas in the disc. 
Hence, estimating $E_\mathrm{infall}$ is not straightforward and requires a careful modelling of the accretion channel under consideration. 

Another gravity-driven mechanism that may sustain turbulence is the gravitational instability of the galactic disc, which is usually studied using the Toomre parameter 
\citep{1964Toomre}. 
For example, \cite{2016Krumholz} tested two models of turbulence, one based on feedback from star formation and the other on gravitational instability. 
They compared both models to measurements of the velocity dispersion of HI and H$\alpha$ lines in both local and distant galaxies, finding that the gravity-driven mode is favoured 
only in star-forming galaxies with high velocity dispersion ($\gtrsim 50$ \kms) and high SFR ($\gtrsim 10 \text{ M}_\odot \text{yr}^{-1}$). 
This is not the case of our galaxies, which have standard velocity dispersions (about 6--15 \kms) and star formation rates (i.e. from SFR $\approx 0.005\text{ M}_\odot \text{yr}^{-1}$ 
for DDO~154 to SFR $\approx 3.2\text{ M}_\odot \text{yr}^{-1}$ for NGC~6946; \citealt{2008Leroy}). 
Moreover, whether these high velocity dispersions are real or affected by observational biases is debated. 
Several authors have shown indeed that, if the effect of beam smearing is properly taken into account, the velocity dispersions in distant galaxies are comparable to or only slightly 
larger than in local galaxies \citep[e.g.][]{2016DiTeodoro,2018DiTeodoro,2018Lelli}. 

In this work, we have shown that the radial profile of the observed energy can be perfectly reproduced by our simple model of energy injection by SN feedback with constant efficiency. 
We find no indication that any additional source of energy is required to sustain turbulence in our sample of galaxies. 
Moreover, the estimates of the energy injected by SNe is the least uncertain considering all the issues related to the other possible driving mechanisms. 
We conclude that SN feedback is likely the most important driver of turbulence in the ISM of nearby galaxies. 

\section{Summary and conclusions}\label{sec:conclusions}
The aim of this work is understanding whether SN feedback can sustain turbulence in the atomic gas and the molecular gas of star-forming galaxies. 
The distribution and kinematics of HI \citep[see][]{2019Bacchini} and CO were derived using emission lines data cubes for a sample of 10 nearby galaxies, allowing us to 
calculate the kinetic energy per unit area as a function of the galactocentric radius.  
We adopted a simple model based on the idea that the gas is in hydrostatic equilibrium and its kinetic energy is given by the sum of two components, namely the turbulent energy and 
the thermal energy. 
Relying on Kolmogorov's framework, we assumed that the turbulent energy is entirely supplied by SN feedback with efficiency $\eta$, corresponding to the fraction of the total energy 
that is transferred to the ISM as kinetic energy. 
The rate of SN explosions per unit area is derived from the observed SFR surface density. 
We also assumed that the driving scale of SN feedback is $L_\mathrm{D}=2 h$, where $h$ is the scale height of the gas in hydrostatic equilibrium. 
The increase of the scale height with the galactocentric radius has a crucial impact on the timescale of turbulence dissipation, which is estimated to be one order of magnitude longer at 
larger radii than in the inner regions of the disc. 
For the atomic gas in particular, we explored two extreme scenarios with either all CNM or all WNM, in which the thermal motions give respectively the minimum and the maximum possible 
contribution to the total energy of the gas, and a more realistic case with a mixture of the two phases. 
We use a Bayesian method to compare our model to a set of observations, aiming to estimate the SN efficiency required to maintain turbulence. 
We found that the radial profiles of the observed energy per unit area in our sample of galaxies are reproduced by the SN feedback model with $\eta$ constant with the galactocentric radius 
at values always below 0.1. 

Our main conclusions are the following:
\begin{enumerate}
 \item At most a few percent of the energy from SN feedback is required to sustain the gas turbulence in our galaxies. 
 We estimate that the median SN efficiency is $\langle \eta_\mathrm{atom} \rangle \approx 0.015$ for the atomic gas and $\langle \eta_\mathrm{mol} \rangle \approx 0.003$ 
 for the molecular gas.
 Therefore, no additional sources of turbulent energy are needed. 
 \item Thermal motions significantly contribute to the observed kinetic energy and velocity dispersion of the atomic gas, especially in the outer and low-SFR regions of galaxies.   
\end{enumerate}
These findings show that low-efficiency SN feedback is sufficiently energetic to be the sole driver of turbulence in local star-forming galaxies. 

The results presented in this work and in \cite{2019Bacchini,2019Bacchini_b} provide empirical indications that SN feedback and star formation are part of the same 
self-regulating cycle \citep[e.g.][]{1985Dopita,2011Ostriker,2020Sun}. 
In this scenario, the SFR per unit volume of a galaxy depends on the volume density of the gas as $\rho_\mathrm{SFR} \propto \rho_\mathrm{gas}^2$. 
The balance between gravity and gas pressure is set by the velocity dispersion of the gas, which includes both thermal and turbulent motions. 
These latter are sustained by SN feedback and therefore depends on the SFR itself. 
In future works, we plan to study this self-regulating cycle and its role in galaxy formation and evolution. 

\section*{Acknowledgements}
CB is grateful to E. di Teodoro for the help and the advice in the analysis of the THINGS data cubes. 
GP acknowledges support by the Swiss National Science Foundation, grant PP00P2\_163824. 
This work was carried out using the publicly available data cubes from HERACLES \citep{2005Leroy}, THINGS \citep{2008Walter}, and HALOGAS \citep{2011Heald}. 
%

\bibliographystyle{aa}
\bibliography{paty.bib}

%

\appendix

\section{Detailed molecular gas kinematics}\label{ap:3DB}
The kinematics of molecular gas was analised using $^{\text{\textsc{3D}}}$\textsc{Barolo} on HERACLES data cubes (see Sec.~\ref{sec:obs_CO}), which have a channel separation of 5.2 \kms 
and a spatial resolution of about 13\arcsec. 
We smoothed the data cubes of NGC~2403 to 27\arcsec, NGC~2976 to 23\arcsec, and NGC~4736 to 18\arcsec, in order to match with the working resolution adopted in \citetalias{2019Bacchini}, 
and to increase the signal-to-noise ratio. 
The inclination and the position angle were taken from \citetalias{2019Bacchini}. 
Small corrections of 2\degr--3\degr~were applied to NGC~5055 and NGC~6946 after an exploratory fit of the data cube. 
Indeed, the values reported in \citetalias{2019Bacchini} were derived from the HI emission, which is more extended than the CO emission, and the inclination and position angle of the 
outer regions of the HI disc may not be the best choice for the CO disc in the presence of a warp. 
The systemic velocities are the same as in \citetalias{2019Bacchini} except for three galaxies, which required a small correction to obtain a better fit the CO data cube: NGC4~736 (+13.3 \kms), 
NGC~5055 (+8.3 \kms), and NGC~6946 (+14.3 \kms). 
The panels in Figs.~\ref{fig:NGC0925}--\ref{fig:NGC6946} provide the main information about the data cubes and the best-fit model of each galaxy\footnote{In all the fits, we used the 
following set of $^{\text{\textsc{3D}}}$\textsc{Barolo} parameters: \texttt{ftype=1}, \texttt{wfunc=2}, \texttt{ltype=1}, \texttt{norm=local}, \texttt{mask=smooth}, \texttt{side=B}.}. 
The description of each panel follows below:
\begin{itemize}
 \item Panel A: 0th moment map of CO emission. 
 The white cross indicates the galaxy center and the white ellipse corresponds to the outermost fitted ring;
 \item Panel B: velocity field or 1st moment map of the data cube. 
 The thick contour shows the systemic velocity and the black circle in the bottom right corner represents the beam of the telescope or the adopted beam after smoothing, as explained above;
 \item Panel C: velocity dispersion map obtained as the 2nd moment map of the data cube. 
 The black bar in the bottom right corner shows the physical scale of the observations;
 \item Panel D: molecular gas surface density as a function of the galactocentric radius $R$ from \cite{2016Frank} (see Sec.~\ref{sec:obs_CO} for details);
 \item Panel E: rotation velocity as a function of $R$;
 \item Panel F: velocity dispersion radial profile. This is not obtained from the 2nd moment map but through the 3D modelling.  
 The dotted horizontal line shows the velocity resolution limit, which is $\approx 0.85 \Delta \upsilon_\mathrm{ch} \approx 4.4$ \kms for Hanning-smoothed data cubes with channel separation 
  $\Delta \upsilon_\mathrm{ch}=5.2$ \kms. 
 The points marked with the red cross (when present) were excluded from this study;
 \item Panel G: inclination of the rings, which was kept constant with $R$ (the dashed grey line shows the mean value);
 \item Panel H: position angle of the rings. The dashed grey line shows the mean value and the red curve (when present) indicates the regularised profile used to obtain the 
 final model (see \citealt{2015Diteodoro} for details);
 \item Panel I: position--velocity diagram along the major axis of the CO disc. 
 The black and the red contours are the iso-density contours of the galaxy and the best-fit model, respectively. 
 The horizontal black dashed line shows the systemic velocity.
 \item Panel J: position--velocity diagram along the minor axis. 
\end{itemize}

\begin{figure*}
\includegraphics[width=2.\columnwidth]{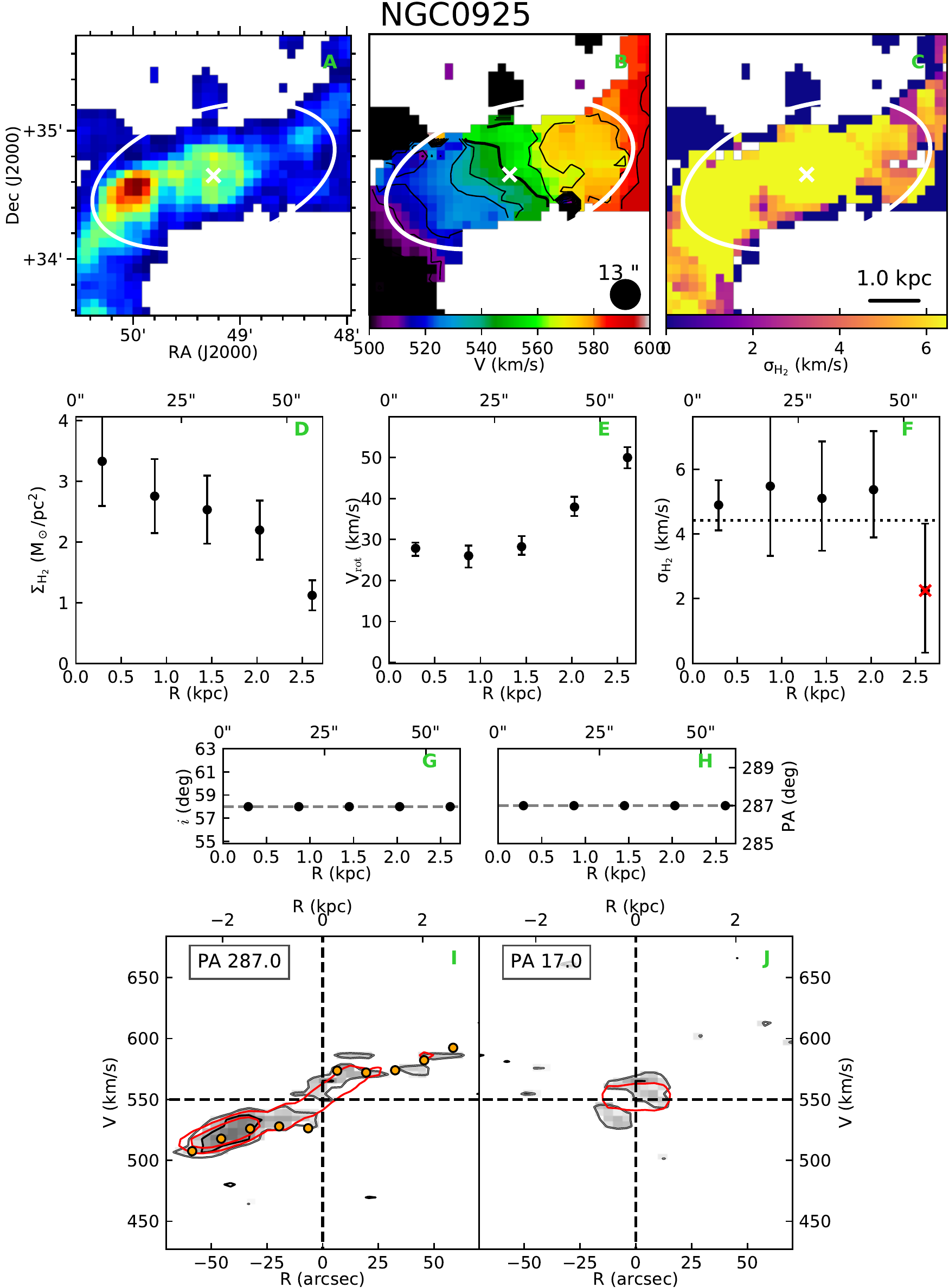}
\caption{}
\label{fig:NGC0925}
\end{figure*}

\begin{figure*}
\includegraphics[width=2.\columnwidth]{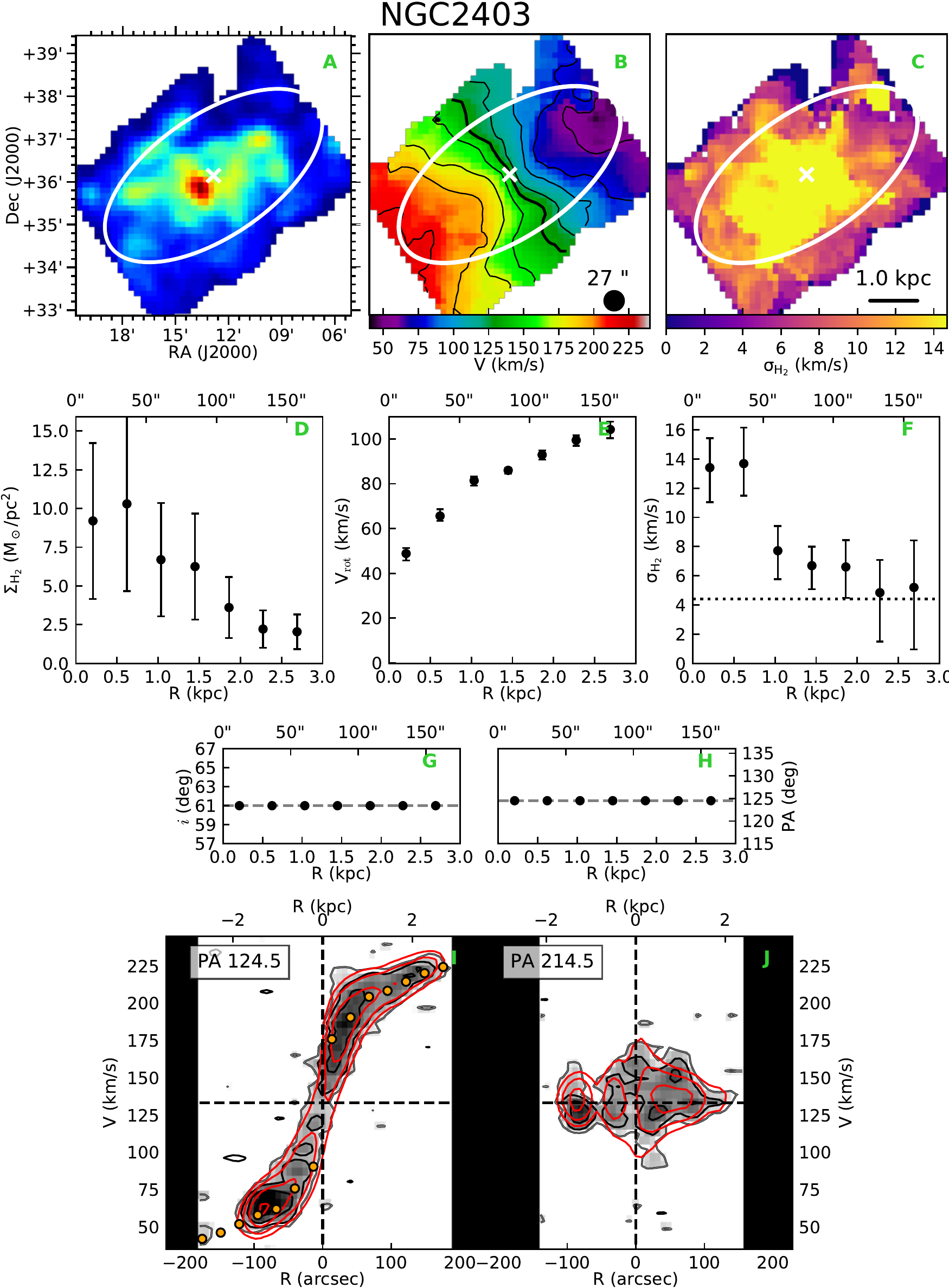}
\caption{}
\label{fig:NGC2403}
\end{figure*}

\begin{figure*}
\includegraphics[width=2.\columnwidth]{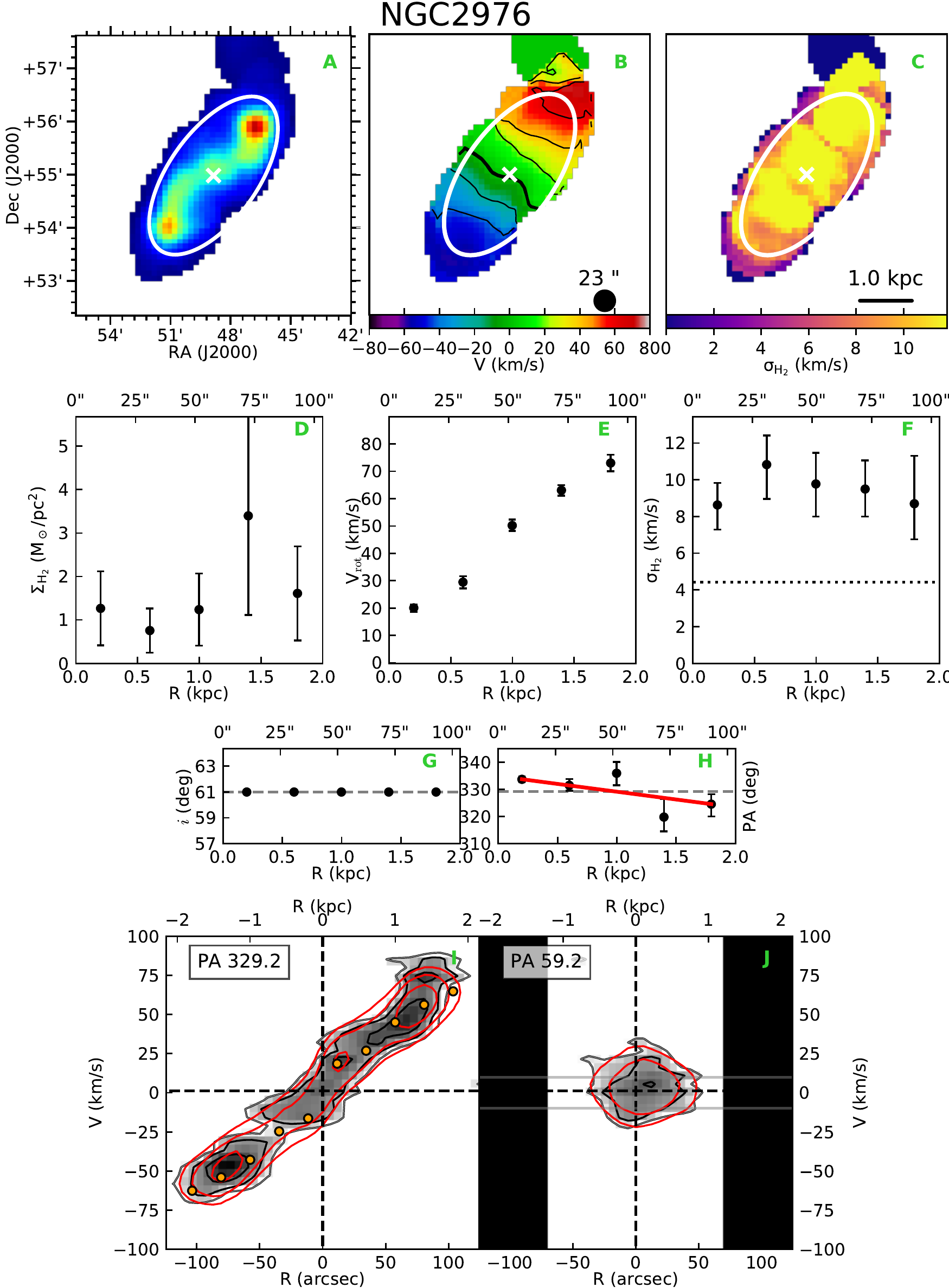}
\caption{}
\label{fig:NGC2976}
\end{figure*}

\begin{figure*}
\includegraphics[width=2.\columnwidth]{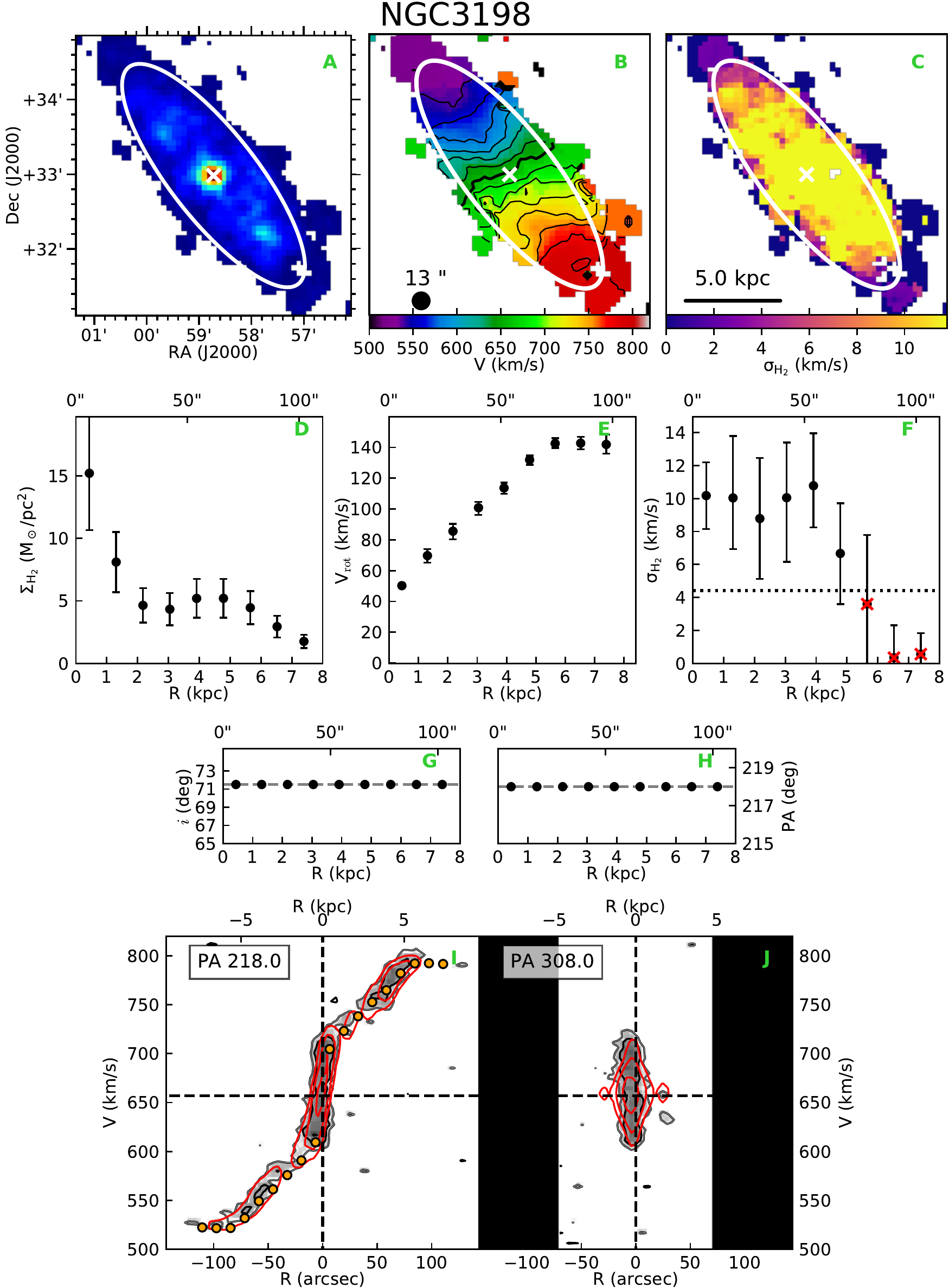}
\caption{}
\label{fig:NGC3198}
\end{figure*}

\begin{figure*}
\includegraphics[width=2.\columnwidth]{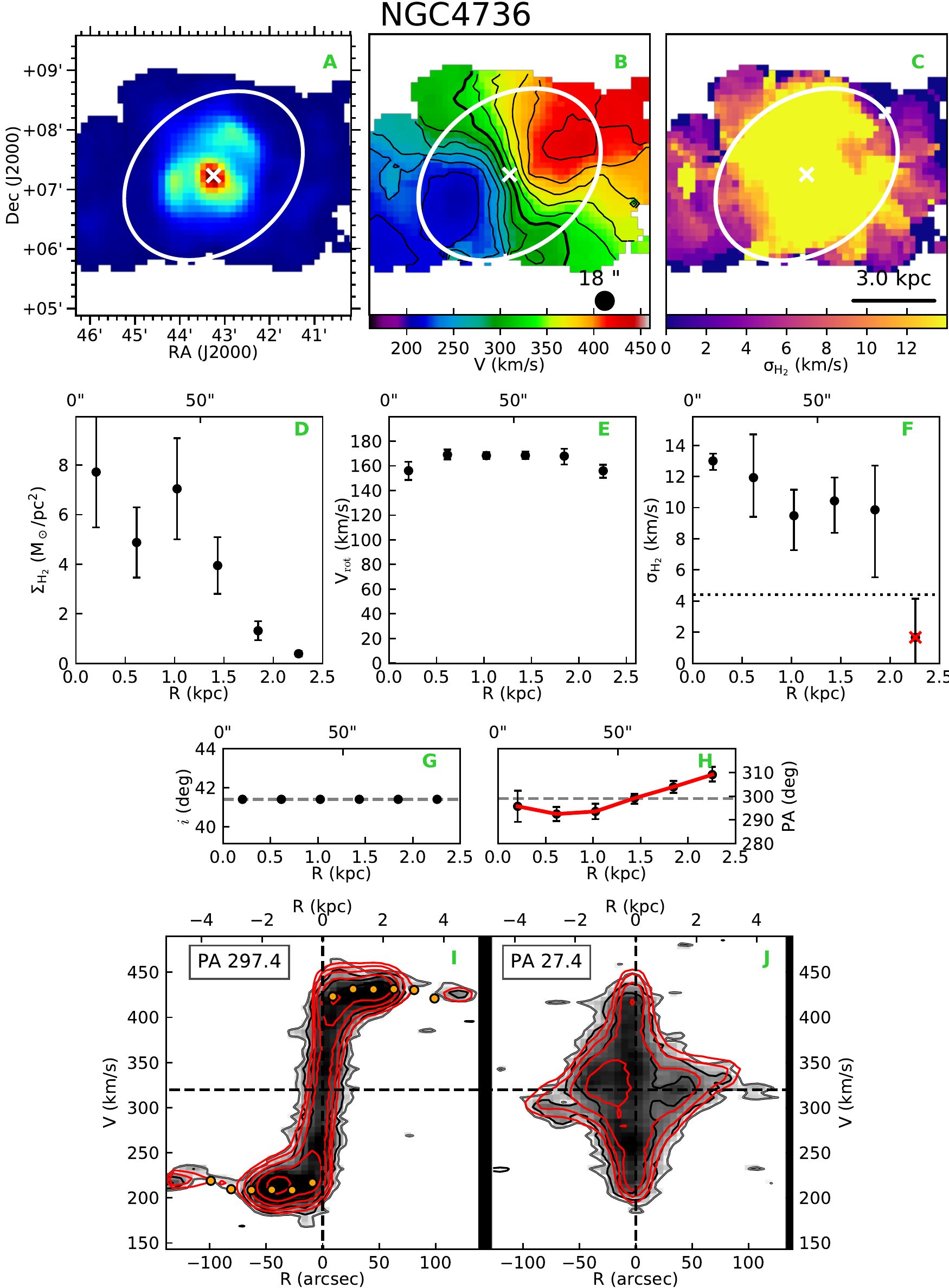}
\caption{}
\label{fig:NGC4736}
\end{figure*}

\begin{figure*}
\includegraphics[width=2.\columnwidth]{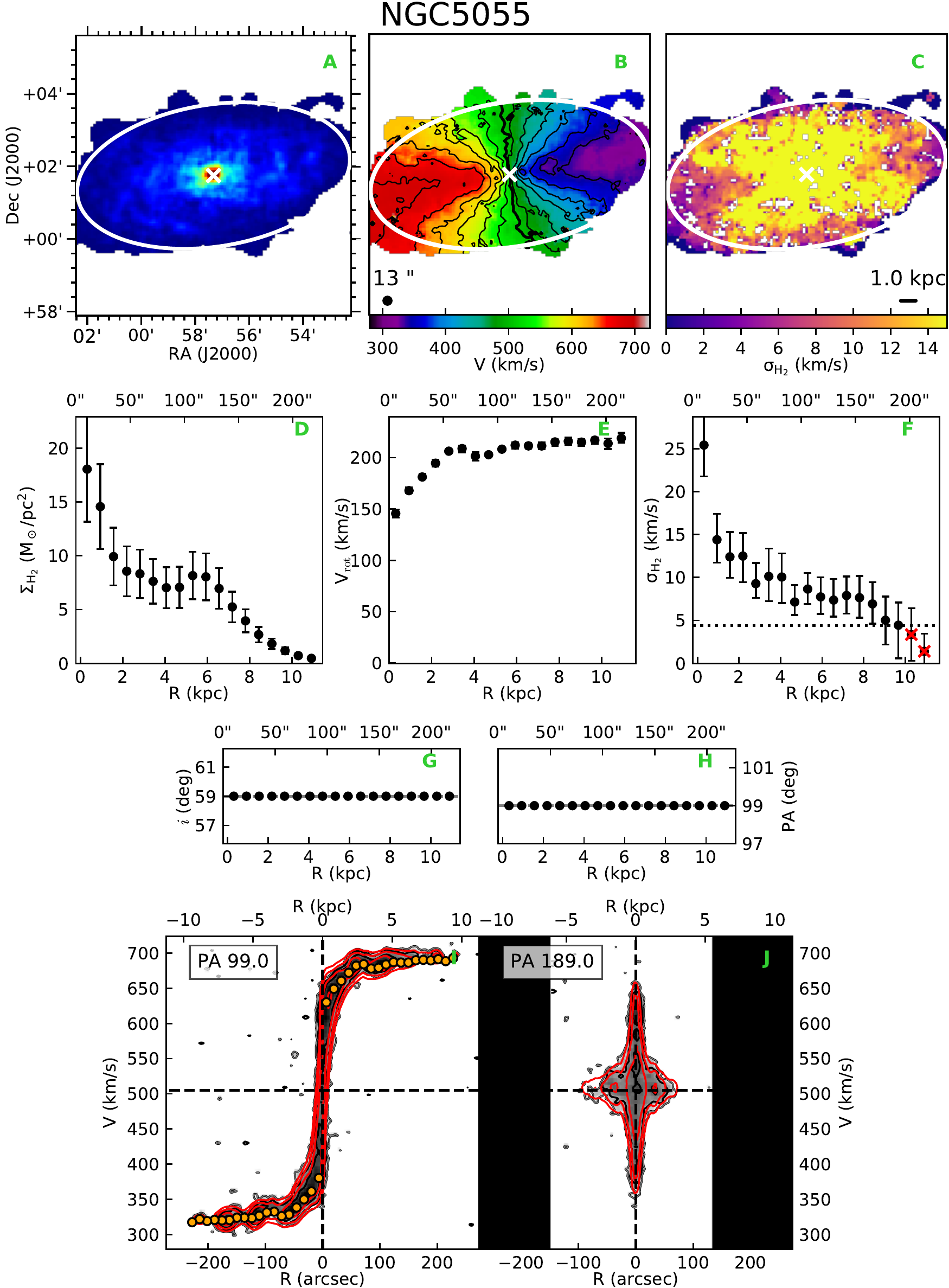}
\caption{}
\label{fig:NGC5055}
\end{figure*}

\begin{figure*}
\includegraphics[width=2.\columnwidth]{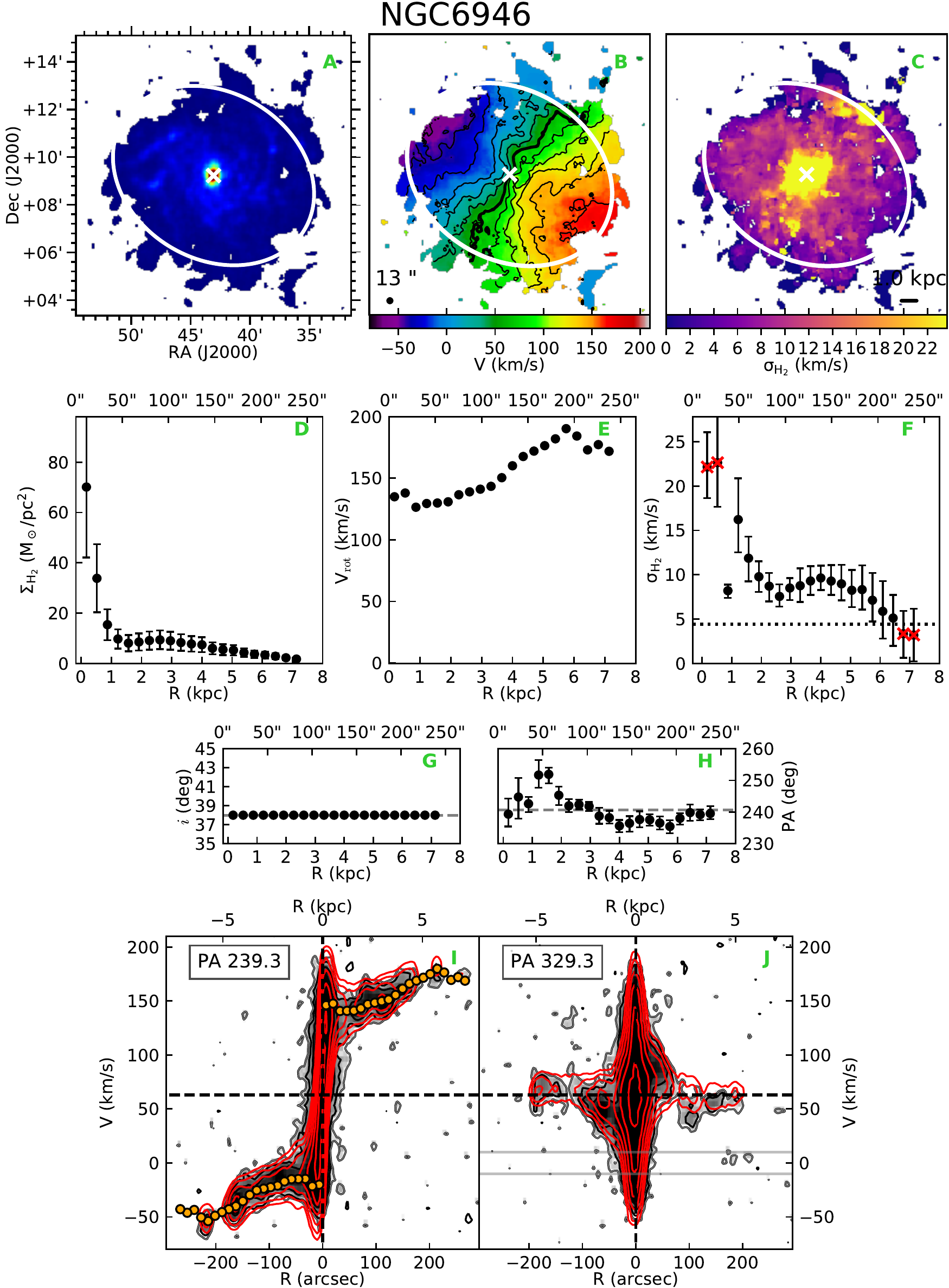}
\caption{}
\label{fig:NGC6946}
\end{figure*}

From panels F in Fig.s~\ref{fig:NGC0925}--\ref{fig:NGC6946}, we can see that the CO velocity dispersion decreases with increasing $R$, going from 10--20 \kms 
in the inner regions to 4--8 \kms at the largest radii, very similar to the HI velocity dispersion profiles obtained by many authors 
(e.g. \citealt{2002Fraternali,2008Boomsma,2009Tamburro,2017Iorio}, and \citetalias{2019Bacchini}). 
However, the distribution of CO is NGC~0925 and NGC~2976 extends only at $R \sim 2$~kpc, hence the decline is not appreciable. 

We excluded some points from the analysis, which are indicated with red crosses in the panel F of each galaxy. 
In particular, the velocity dispersion measured at the outermost radii of most galaxies (i.e. NGC~0925, NGC~3198, NGC~4736, NGC~5055, and NGC~6946) is excluded, as it is lower than the data 
cubes velocity resolution. 
From the position-velocity diagrams of NGC~4736 (Fig.~\ref{fig:NGC4736}) and NGC~6946 (Fig.~\ref{fig:NGC6946}), we can see that the red contours of $^{\text{\textsc{3D}}}$\textsc{Barolo} 
model are thinner than the black ones of the galaxy emission in the regions within about 1~kpc from the center. 
This is due to the presence of a bar, which drives strong non-circular motions and produces wiggles in the velocity field \citep[see also][]{1995Moellenhoff,2008Boomsma}. 
These kinematic feature cannot be reproduced by the tilted-ring model, as it assumes circular motions. 
NGC~0925 is also barred galaxy \citep[e.g.][]{1998Pisano}, but in this case the feature is not evident from the position-velocity diagram (Fig.~\ref{fig:NGC0925}) 
due to the low signal-to-noise ratio of the data. 
However, the odd shape of the rotation curve suggests that there could be non-circular motions in the region within $R\sim1$~kpc, which cannot be caught by the tilted-ring model. 
We note that the position angle of NGC~4736 (see panel B in Fig.~\ref{fig:NGC4736}) is appropriate for the outer disc, while the velocity field in inner regions suggests a lower value. 
In the case of NGC~6946, we adopted an \textit{ad hoc} approach to improve the model. 
We first ran $^{\text{\textsc{3D}}}$\textsc{Barolo} setting the velocity dispersion at 4 \kms, in order to retrieve a good-quality rotation curve. 
Then, we performed a second run fixing the rotation velocities of the rings to those obtained previously, and fitting the velocity dispersion. 
After inspecting the position-velocity diagram of NGC~6946 (Fig.~\ref{fig:NGC6946}), we decided however to remove the first two inner points, as the velocity dispersion is clearly overestimated. 

Figure~\ref{fig:COHI_ratio} shows, for each galaxy in the sample, the radial profile of the ratio of $\sigma_\mathrm{CO}$ to $\sigma_\mathrm{HI}$. 
The grey area represents the 84th and the 16th percentile of all the points, which enclose also the median value indicated by the black dash-dotted line 
$\langle \sigma_\mathrm{CO}/ \sigma_\mathrm{HI}\rangle = 0.6 \pm 0.2$. 
This value is in agreement with previous works \citep[e.g.][]{2016Mogotsi,2017Marasco,2019Koch} and additionally confirms the assumption made in \citetalias{2019Bacchini}
We note however that the velocity resolution of HERACLES data cubes is not optimal to study the molecular gas velocity dispersion, which may be even lower than 4.4 \kms at 
large radii. 
Hence, the median value that we found could be slightly overestimated. 
\begin{figure}
\includegraphics[width=1.\columnwidth]{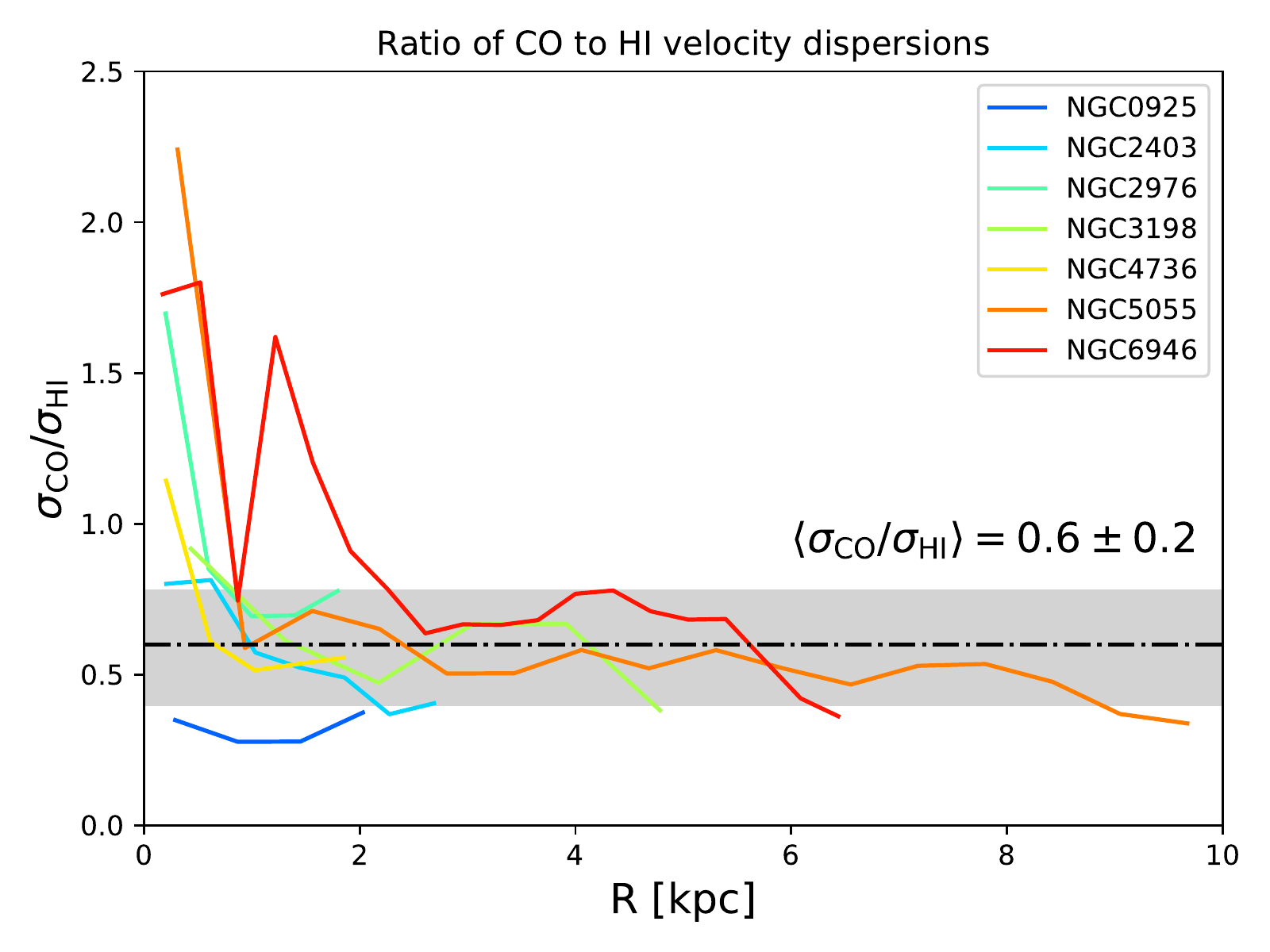}
\caption{Ratio of the CO-to-HI velocity dispersion as a function of the galactocentric radius for the galaxies in our sample with CO detected. 
The median and the 84th and the 16th percentiles are represented by the dot-dashed black line and the grey area, respectively, and their values are reported above.}
\label{fig:COHI_ratio}
\end{figure}

\section{Hierarchical Bayesian inference}\label{ap:HBay}
In this section, we describe the formalism of the Bayesian method presented in Sec.~\ref{sec:method_HB}. 
We estimate, for each galaxy in our sample, the posterior distribution of $\eta$, assuming that it is constant with the galactocentric radius $R$. 
For a single galaxy divided in $N$ annuli, the data are $\mathcal{D}= \{ \mathbf{d}_i \}_{i=1}^N$. 
Let us define $R_\mathrm{SFR}$ as the outermost radius where the SFR surface density is measured. 
Hence, we have the upper limit on $\Sigma_\mathrm{SFR}$ where $R_i > R_\mathrm{SFR}$, which is the case of DDO~154, NGC~2403, NGC~3198, and NGC~6946. 
For the other galaxies instead, $R_i \leq R_\mathrm{SFR}$ at any radius. 
The observed quantities at a certain radius $R_i$ are therefore
\begin{equation}
 \mathbf{d}_i = \left( \Sigma_{\mathrm{SFR}_i} , \sigma_{_i} , h_\mathrm{HI_i} , \Sigma_{\mathrm{HI}_i} \right) \, \text{ where } \, R_i \leq R_\mathrm{SFR} \, ,
\end{equation}
and 
\begin{equation}
 \mathbf{d}_i = \left( \sigma_{_i} , h_\mathrm{HI_i} , \Sigma_{\mathrm{HI}_i} \right) \, \text{ where } \, R_i > R_\mathrm{SFR} \, . 
\end{equation}
The associated uncertainties are
\begin{equation}
 \Delta \mathbf{d}_i = \left( \Delta \Sigma_{\mathrm{SFR}_i} , \Delta \sigma_{_i} , \Delta h_\mathrm{HI_i} , \Delta \Sigma_{\mathrm{HI}_i} \right) \, 
 \text{ where } \, R_i \leq R_\mathrm{SFR} \, ,
\end{equation}
and 
\begin{equation}
 \Delta \mathbf{d}_i = \left( \Delta \sigma_{_i} , \Delta h_\mathrm{HI_i} , \Delta \Sigma_{\mathrm{HI}_i} \right) \, \text{ where } \, R_i > R_\mathrm{SFR} \, . 
\end{equation}
As a preliminary step, the observed quantities for each galaxy were normalised to their median value. 
The elements of $\mathbf{d}_i$ are considered realisation of normal distributions centered on true values (indicated with the apex $^T$) 
and with standard deviation given by the uncertainties in $\Delta \mathbf{d}_i$. 
In order to avoid negative true values, we used log-normal distributions as priors (i.e. $\upsilon_\mathrm{turb_i}^T$, $\Sigma_{\mathrm{SFR}_i}^T$, and $h_\mathrm{HI_i}^T$). 
Each log-normal distribution is defined by two parameters: the central value ($\alpha_\mathrm{\upsilon_\mathrm{turb_i}}$, $\alpha_{\Sigma_{\mathrm{SFR}_i}}$, and $\alpha_{h_\mathrm{HI_i}}$) 
and the standard deviation ($\beta_\mathrm{\upsilon_\mathrm{turb_i}}$, $\beta_{\Sigma_{\mathrm{SFR}_i}}$, and $\beta_{h_\mathrm{HI_i}}$), whose hyper-priors are a normal distribution 
and a Gamma distribution, respectively (see Table~\ref{tab:hyper}). 
The thermal velocity is a normal distribution with centroid $\mu_{\upsilon_\mathrm{th}}$ and standard deviation $\sigma_{\upsilon_\mathrm{th}}$ of 0.06 \kms and 0.005 \kms for CO, 
1 \kms and 0.4 \kms for the cold HI, and 8.1 \kms and 0.5 \kms for the warm HI. 
The true velocity dispersion is calculated with Eq.~\ref{eq:sigma_obs_def}. 
Following \S 5 in \cite{gelman_book}, the prior on $\eta$ is defined by a beta-distribution with exponents $\gamma_\eta= \epsilon \zeta$ and $\delta_\eta=\zeta(1-\epsilon)$, 
whose hyper-parameters $\epsilon$ and $\zeta$ are distributed as reported in Table~\ref{tab:hyper}. 
Hence, the prior on $\eta$ is a continuous distribution between 0 and 1, which is equivalent to a uniform distribution in the particular case $\gamma_\eta=\delta_\eta=1$. 
In addition, this prior of $\eta$ allows us to explore, at the same time, other distributions than the uniform one, adding a further level of generalisation. 
{\rowcolors{5}{white}{white}
	\renewcommand{\arraystretch}{1.5}
	\begin{table}
		\centering
		\caption{Hyper-priors probability distributions: 
			$\mathsf{Normal}$ indicates a normal distribution with central value $\mu_\alpha$ and standard deviation $\sigma_\alpha$;
			$\mathsf{Exp}$ is an exponential distribution with scale $\lambda_\beta$; 
			$\mathsf{Uniform}$ is a continuous uniform distribution between a minimum $l_\epsilon$ and a maximum $u_\epsilon$; 
			$\mathsf{Gamma}$ is a gamma distribution with shape parameter $\gamma_\zeta$ and rate parameter $\delta_\zeta$ \citep[see][]{gelman_book}. 
		}
		\label{tab:hyper}
		\begin{tabular}{c|c}
			\hline\hline
			Hyper-priors													& Probability distribution				\\
			\hline
			$\alpha_\mathrm{\upsilon_\mathrm{turb_i}}$, $\alpha_{\Sigma_{\mathrm{SFR}_i}}$, $\alpha_{h_\mathrm{HI_i}}$	& $\mathsf{Normal} (\mu_\alpha=0,\sigma_\alpha=2)$	\\
			$\beta_\mathrm{\upsilon_\mathrm{turb_i}}$, $\beta_{\Sigma_{\mathrm{SFR}_i}}$, $\beta_{h_\mathrm{HI_i}}$		& $\mathsf{Exp}(\lambda_\beta=1)$			\\
			$\epsilon$													& $\mathsf{Uniform}(l_\epsilon=0,u_\epsilon=1)$		\\
			$\zeta$														& $\mathsf{Gamma}(\gamma_\zeta=1,\delta_\zeta=1/20)$	\\
			\hline
		\end{tabular}
	\end{table}

Where $R_i \leq R_\mathrm{SFR}$, the likelihood in Eq.~\ref{eq:bayes_th_HB} is written as
\begin{equation}\label{eq:like}
\begin{split}
 p \left(\mathcal{D} | \Theta \right) & = 
					\prod_{i=1}^N p \left(\mathbf{d}_i | \Theta \right)  
				      = \prod_{i=1}^N p \left( \Sigma_{\mathrm{SFR}_i} | \Sigma_{\mathrm{SFR}_i}^T, \Delta \Sigma_{\mathrm{SFR}_i} \right) \\
				      & p \left( h_\mathrm{HI_i} | h_\mathrm{HI_i}^T, \Delta h_\mathrm{HI_i} \right) 
				      p \left( \sigma_{_i} | \Delta \sigma_{_i} , \upsilon_\mathrm{turb_i}^T, \upsilon_\mathrm{th} \right) \\
				      & p \left( \Sigma_{\mathrm{HI}_i} | \Delta \Sigma_{\mathrm{HI}_i} , \upsilon_\mathrm{turb_i}^T, \Sigma_{\mathrm{SFR}_i}^T, 
				      h_\mathrm{HI_i}^T \right) \, ,
  \end{split} 
\end{equation}
while, where $R_i > R_\mathrm{SFR}$, it is 
\begin{equation}\label{eq:like_nan}
\begin{split}
 p \left(\mathcal{D} | \Theta \right) & = 
					\prod_{i=1}^N p \left(\mathbf{d}_i | \Theta \right)  
				      = \prod_{i=1}^N p \left( h_\mathrm{HI_i} | h_\mathrm{HI_i}^T, \Delta h_\mathrm{HI_i} \right) \\
				      & p \left( \sigma_{_i} | \Delta \sigma_{_i} , \upsilon_\mathrm{turb_i}^T, \upsilon_\mathrm{th} \right) 
				      p \left( \Sigma_{\mathrm{HI}_i} | \Delta \Sigma_{\mathrm{HI}_i} , \upsilon_\mathrm{turb_i}^T, \Sigma_{\mathrm{SFR}_i}^T, 
				      h_\mathrm{HI_i}^T \right) \, . 
  \end{split} 
\end{equation}
In Eq.~\ref{eq:like} and Eq.~\ref{eq:like_nan}, the probability distributions of $\Sigma_{\mathrm{SFR}_i}$, $\sigma_{_i}$, $h_\mathrm{HI_i}$, and $\Sigma_{\mathrm{HI}_i}$ 
are normal distributions. 
Then, the probability distribution of the priors in Eq.~\ref{eq:bayes_th_HB}, where $R_i \leq R_\mathrm{SFR}$, is
\begin{equation}\label{eq:priors}
\begin{split}
 \pi \left( \Theta | \Phi \right) & = \pi \left( \eta | \gamma_\eta , \delta_\eta \right) 
				      \prod_{i=1}^N 
				      \pi \left( \upsilon_\mathrm{turb_i}^T | \alpha_{\upsilon_\mathrm{turb_i}}, \beta_{\upsilon_\mathrm{turb_i}} \right) \\
				    & \pi \left( \Sigma_{\mathrm{SFR}_i}^T | \alpha_{\Sigma_{\mathrm{SFR}_i}}, \beta_{\Sigma_{\mathrm{SFR}_i}} \right) 
				      \pi \left( h_\mathrm{HI_i}^T | \alpha_{h_\mathrm{HI_i}}, \beta_{h_\mathrm{HI_i}} \right) \\
				    & \pi \left( \sigma_{_i}^T |  \upsilon_\mathrm{turb_i}^T, \upsilon_\mathrm{th} \right) \, ,
\end{split} 
\end{equation}
where the probability distributions of $\Sigma_{\mathrm{SFR}_i}^T$, $\sigma_{_i}^T$, $h_\mathrm{HI_i}^T$, and $\Sigma_{\mathrm{HI}_i}^T$ are log-normal distributions. 
Where $R_i > R_\mathrm{SFR}$, we have instead
\begin{equation}\label{eq:priors_uplim}
\begin{split}
 \pi \left( \Theta | \Phi \right) & = \pi \left( \eta | \gamma_\eta , \delta_\eta \right) 
				      \prod_{i=1}^N 
				      \pi \left( \upsilon_\mathrm{turb_i}^T | \alpha_{\upsilon_\mathrm{turb_i}}, \beta_{\upsilon_\mathrm{turb_i}} \right) \\
				    & \pi \left( \Sigma_{\mathrm{SFR}_i}^T | 0, u_{\Sigma_\mathrm{SFR}} \right) 
				      \pi \left( h_\mathrm{HI_i}^T | \alpha_{h_\mathrm{HI_i}}, \beta_{h_\mathrm{HI_i}} \right) \\
				    & \pi \left( \sigma_{_i}^T |  \upsilon_\mathrm{turb_i}^T, \upsilon_\mathrm{th} \right) \, ,
\end{split} 
\end{equation}
and the probability distributions of $\Sigma_{\mathrm{SFR}_i}^T$ is a uniform distribution between 0 and the upper limit $u_{\Sigma_\mathrm{SFR}}$. 

In the case of the two-phase HI, the probability distribution of the thermal velocity is a uniform distribution between 
$l_\mathrm{\upsilon_\mathrm{th}} = \upsilon_\mathrm{th,c} = 1 \text{ \kms}$, which is the value for the CNM, and 
$u_\mathrm{\upsilon_\mathrm{th}} = \upsilon_\mathrm{th,w} = 8.1 \text{ \kms}$, which is the value for the WNM. 

We note that our best models (green bands in Figs.~\ref{fig:allgals_results_SNe_HI_cnm}-\ref{fig:E_radial_H2cnm}) are not directly obtained from $E_\mathrm{obs}$ (black points in Figs.~\ref{fig:allgals_results_SNe_HI_cnm}-\ref{fig:E_radial_H2cnm}), which is derived through Eq.~\ref{eq:E_obs_def} using the observed velocity dispersion and surface density (the error bars are calculated with the uncertainty propagation rules). As explained in this section and in Sec.~\ref{sec:method_HB}, our procedure is based on the priors on the observable quantities (Eq.~\ref{eq:priors} and Eq.~\ref{eq:priors_uplim}). We can define a "true" observed energy ($E_\mathrm{obs}^T$) and derive its posterior distribution through Eq.~\ref{eq:E_obs_def} using the priors on $\Sigma$ and $\sigma^T$. The median and 1$\sigma$ error of $E_\mathrm{obs}^T$ are not necessarily the same as the value and the error bar of $E_\mathrm{obs}$. This latter is directly derived from the observed quantities, while $E_\mathrm{obs}^T$ is "theoretical". Hence, it is formally correct to compare $E_\mathrm{mod}$ with $E_\mathrm{obs}$. This specification applies to all the figures shown in Sec.~\ref{sec:results}.

\end{document}